\documentclass[proof]{pasj00}
\draft 
\begin{document}


\SetRunningHead{Goto et al.}{Morphological Butcher-Oemler Effect}
\title{Morphological Butcher-Oemler effect\\ in the SDSS Cut~\&~Enhance Galaxy Cluster Catalog}

%

\author{%
   Tomotsugu \textsc{Goto}\altaffilmark{1,2,3},
   Sadanori \textsc{Okamura}\altaffilmark{2},
   Masafumi \textsc{Yagi}\altaffilmark{4},
   Ravi K. \textsc{Sheth}\altaffilmark{5},\\
   Neta A. \textsc{Bahcall}\altaffilmark{6},
   Shane A. \textsc{Zabel}\altaffilmark{7},
   Michael S. \textsc{Crouch}\altaffilmark{7},
       Maki \textsc{Sekiguchi}\altaffilmark{1} ,\\
   James \textsc{Annis}\altaffilmark{8},
   Mariangela \textsc{Bernardi}\altaffilmark{7},
   Shang-Shan \textsc{Chong}\altaffilmark{7},
   Percy L. \textsc{G{\' o}mez}\altaffilmark{7},\\
   Sarah \textsc{Hansen}\altaffilmark{9},
   Rita S. J. \textsc{Kim}\altaffilmark{10},
    Adam \textsc{Knudson}\altaffilmark{7},
   Timothy A. \textsc{Mckay}\altaffilmark{9},\\
   and Christopher J. \textsc{Miller}\altaffilmark{7}
}   
 \altaffiltext{1}{Institute for Cosmic Ray Research, University of
   Tokyo,\\ Kashiwanoha, Kashiwa, Chiba 277-0882, Japan}
 \altaffiltext{2}{Department of Astronomy and Research Center for the
   Early Universe,\\ School of Science, University of Tokyo, Tokyo
   113-0033, Japan}
 \altaffiltext{3}{yohnis@icrr.u-tokyo.ac.jp}
\altaffiltext{4}{
National Astronomical Observatory, 2-21-1 Osawa, Mitaka, Tokyo 181-8588,
Japan.}
 \altaffiltext{5}{ Department of Physics and Astronomy
 University of Pittsburgh\\
 3941 O'Hara Street Pittsburgh, PA 15260  }
 \altaffiltext{6}{Princeton University Observatory, Princeton, NJ 08544,
 USA}
 \altaffiltext{7}{Department of Physics, Carnegie Mellon University, \\
 5000 Forbes Avenue, Pittsburgh, PA 15213-3890, USA}
 \altaffiltext{8}{Fermi National Accelerator Laboratory, P.O. Box 500,
   Batavia, IL 60510, USA }
 \altaffiltext{9}{University of Michigan, Department of Physics,\\ 500
   East University, Ann Arbor, MI 48109, USA}
 \altaffiltext{10}{Department of Physics and Astronomy, The Johns Hopkins
   University,\\ 3400 North Charles Street, Baltimore, MD 21218-2686, USA}


%


\maketitle

\begin{abstract}\label{abstract}

  We investigate the evolution of the 
 fractions of late type
 cluster galaxies as a function of redshift, using one of the largest,
 most uniform  cluster samples available.  The sample consists of 514 clusters of galaxies in the range
 0.02$\leq$z$\leq$0.3 from the Sloan Digital Sky Survey Cut \& Enhance
 galaxy cluster catalog.
 This catalog was created using a single
 automated cluster finding algorithm applied to uniform data from a single
 telescope, with accurate CCD photometry, thus, 
 minimizing selection biases.
 We use four independent methods to analyze the evolution of the late
 type galaxy fraction.
 Specifically, we select late type galaxies based on: restframe
 $g-r$ color, $u-r$ color, galaxy profile fitting and concentration
 index. 
 The first criterion corresponds to the one used in
 the classical Butcher-Oemler analyses. The last two criteria are more sensitive to the
 morphological type of the galaxies. In all the four cases, we find an
 increase in the fraction of late type galaxies with increasing
 redshift, significant at the 99.9\% level.
 The results confirm that cluster galaxies do
 change colors with redshift (the Butcher-Oemler effect) and, in addition,
 they change their morphology to later-type toward higher redshift --- indicating a 
 morphological equivalent of the Butcher-Oemler effect. 
 We also find a  tendency of richer clusters to have lower fractions of late type
 galaxies. The trend is consistent with a ram pressure
 stripping model, 
  where galaxies in richer clusters are affected by stronger ram
 pressure due to higher temperature of clusters. 
%

 
 

\end{abstract}

\section{Introduction}\label{intro}
  
 The Butcher-Oemler effect was first reported by Butcher \& Oemler
 (1978, 1984) as an increase in the fraction of blue galaxies ($f_b$) toward
 higher redshift in 33 galaxy clusters  over the redshift range
 0$<$z$<$0.54.
 Butcher and Oemler's work made a strong impact since it showed direct evidence for the
 evolution of cluster galaxies. Much work regarding the nature of
 these blue galaxies followed. Rakos \& Schombert (1995) found that the
 fraction of blue galaxies increases from 20\% at z=0.4 to 80\% at
 z=0.9, suggesting that the evolution in clusters is even stronger than
 previously thought. Margoniner \& De Carvalho (2000) studied 48
 clusters in the redshift range of 0.03$<$z$<$0.38 and detected a strong
 Butcher-Oemler effect consistent with  that of Rakos \& Schombert (1995). 
 Despite the trend with redshift,  almost all previous work has reported 
 a wide range of blue fraction values. 
 In particular, in a large sample of 295 Abell clusters, Margoniner et al. (2001)
 not only confirmed the existence of the Butcher-Oemler effect,
 but also found the blue fraction depends on cluster richness.

 Although the detection of the Butcher-Oemler effect has been claimed in
 various studies, 
 there have been some suggestions of strong selection biases in the cluster samples.
 Newberry, Kirshner \& Boroson (1988) measured velocity
 dispersions and surface densities of galaxies in clusters and found a marked difference between local clusters and intermediate
 redshift clusters. More recently, Andreon \& Ettori (1999)
 measured X-ray surface brightness profiles, sizes, and luminosities of the
 Butcher-Oemler sample of clusters and concluded that the sample is not
 uniform. The selection bias, thus, could mimic evolutionary
 effects. Smail et al. (1998) used 10 X-ray bright clusters in the
 redshift range of 0.22$\leq$z$\leq$0.28 and found
 that the clusters have only a small fraction of blue galaxies.  The
 Butcher-Oemler effect was not observed with their sample. 
 Similarly, galaxies in radio selected groups are not significantly
 bluer at higher redshifts (Allington-Smith et al. 1993). 
 Garilli et al. (1996) observed 67 Abell and X-ray selected clusters and
 found no detectable Butcher-Oemler effect until z=0.2. Fairley et
 al. (2002) studied eight X-ray selected galaxy clusters and found no
 correlation of blue fraction with redshift.   Rakos \& Schombert (1995)'s
 sample were selected from the catalog compiled by Gunn, Hoessel \& Oke
 (1986) using photographic plates taken only in two color bands.
 The sample
 thus have a possible bias against red, high redshift clusters.
 In addition to the possible sample selection biases, with the exception of Margoniner et
 al. (2001), the number of clusters in the previous works was small, consisting of a few to dozens of clusters. Therefore the statistical
 uncertainty was large. Many authors also noted
 that cluster-to-cluster variation of the fraction of blue galaxies is considerable.
 The need for a larger, more uniform sample of clusters has been evident. 

 There have been various attempts to find another
 physical mechanism causing the large scatter which has been seen in
 almost all previous work.
 Wang \& Ulmer (1997) claimed the existence of a correlation between the blue fraction
 and the ellipticity of the cluster X-ray emissions in their sample of
 clusters at 0.15$\leq$z$\leq$0.6. Metevier, Romer \& Ulmer (2000) showed that two clusters with a bimodal X-ray surface brightness profile have an unusually high blue fraction
 value and thus do not follow the typical Butcher-Oemler relation. They
 claimed that the Butcher-Oemler effect is an environmental phenomenon
 as well as an evolutionary phenomenon.  Margoniner et al. (2001) found
 a richness dependence in the sense that richer
 clusters have smaller blue fractions.  They claimed that this
 richness dependence causes a large scatter in the blue fraction--redshift
 diagram. Therefore, it is of extreme interest to explore an origin
 of the scatter in the blue fraction despite the redshift trend.

 At the same time,  various studies using morphological information have
 reported a similar evolution effect in cluster galaxies. 
 Dressler et al. (1997) studied 10 clusters at
 0.37$<$z$<$0.56 and found a steep increase in the S0 fraction toward $lower$
 redshift, compared to nearby clusters studied earlier (Dressler 1980). 
 Couch et
 al. (1994,1998) studied three clusters at z=0.31 and found their S0 fraction to be consistent with the
 trend observed by Dressler et
 al. (1997). Fasano et al. (2000) observed nine clusters at intermediate
 redshifts (0.09$<$z$<$0.26) and also found an increase in the S0 fraction
 toward lower redshift. 
 It has been proposed that the increase in the S0 fraction is caused by the
 transformation of spiral galaxies into S0 galaxies through a process
 yet unknown. These studies, however, need to be pursued further, considering that most
 of the previous work was based on morphological galaxy
 classification by eye. Although it is an excellent tool to
 classify galaxies, manual classification could potentially have unknown
 biases. (A detailed comparison of human classifiers can be found in
 Lahav et al. 1995). 
 A machine based, automated classification 
 would better control biases and would allow a reliable determination of
 the completeness and false positive rate. A further reason to
 investigate the evolution of cluster galaxies is the sample size. 
 The morphological fraction studies of clusters reported so
 far are based on only dozens of clusters. Furthermore, the clusters
 themselves have intrinsic variety in the fraction of blue/spiral
 galaxies as reported by various Butcher-Oemler and morphological analyses
 listed above.  Since several authors have  suggested that the fraction of
 blue galaxies depends on cluster richness, it is important
 to use a uniform cluster sample, preferably selected by an automated
 method with a well known selection function.

 Various theoretical models have been proposed to explain the Butcher-Oemler
 effect and the increase of the S0 fraction. These models include ram pressure stripping
 of gas (Spitzer \& Baade 1951, Gunn \& Gott 1972, Farouki \& Shapiro
 1980; Kent 1981, Fujita 1998; Abadi, Moore \& Bower 1999, Fujita \& Nagashima 1999,
 Quilis, Moore \& Bower 
 2000), galaxy infall (Bothun \& Dressler 1986, Abraham et al. 1996a,
 Ellingson et al. 2001), 
 galaxy harassment (Moore et al. 1996, 1999), cluster tidal forces (Byrd
 \& Valtonen 1990, Valluri 1993),  enhanced star formation (Dressler \&
 Gunn 1992),
 and 
  removal and consumption of the gas (Larson,
 Tinsley \& Caldwell 1980, Balogh et al. 2001, Bekki et al. 2002). It is, however, yet
 unknown exactly what processes play major roles in changing the color and
 morphology of cluster galaxies. 
 To derive a clear model explaining 
 the evolution of cluster galaxies, it is important to clarify both the
 Butcher-Oemler effect and the S0 increase, using a large and uniform
 cluster sample in conjunction with a machine based morphological classification. 

 With the advent of the Sloan Digital Sky Survey (SDSS; York et al. 2000),
 which is an imaging and spectroscopic survey of 10,000 deg$^2$ of the
 sky, we now have the opportunity to overcome these limitations. The SDSS Cut
 \& Enhance galaxy cluster catalog (Goto et al. 2002a) provides a large
 uniform cluster catalog with a well defined selection function.
 The CCD-based, accurate photometry of the SDSS (Fukugita
 et al. 1996, Hogg et al. 2002, Smith et al. 2002) and the wide coverage of the SDSS
 on the sky allow accurate estimation of blue fraction with robust
 local background subtraction. Although the SDSS is a ground based
 observation, the state-of-the-art reduction software and the accuracy
 of CCD data make it possible to derive morphological classification in an
 automated way  (Lupton et al. 2001, 2002). By using the SDSS data
 set, we are able to study one of the largest samples to date --- 514 clusters --- to
 the depth of $M_r$=-19.44 ($h$=0.75).  

 The purpose of this paper is as follows. We aim to confirm or disprove the existence
  of the Butcher-Oemler effect using one of the largest, most 
 uniform cluster samples. At the same time, we
 hope to shed light on the morphological properties of the Butcher-Oemler
 galaxies using morphological parameters derived from the SDSS data. 
 Finally we investigate the origin of the scatter in the galaxy type
 fraction versus redshift
 relation, in hope of gaining some understanding
 about the physical processes responsible for the scatter. 

 Since several people reported that field galaxies also evolve both
 morphologically (Schade et al. 1996;  Brinchmann et al. 1998;Lilly et
 al. 1998; Kajisawa et al. 2001;Abraham et al. 2001;) and
 spectroscopically 
 (Madau et al. 1996; Lilly et al.1996; Hammer et al. 1997; Treyer et al. 1998;  Cowie et al. 1999; Sullivan et al. 2000; Wilson et al. 2002),   it is of extreme
 importance to compare the evolution of cluster galaxies with that of field galaxies to
 specify a responsible physical mechanism. 
 It is  possible that the Butcher-Oemler effect and morphological
 transition of cluster galaxies are more commonly happening including
 the field region of the universe, thus a cluster specific mechanism is
 not responsible for the evolution of galaxies. However, since the SDSS spectroscopic data
 are not deep enough to probe cosmologically interesting time scale, we
 leave it to future work. 
  
  The paper is organized as follows: In
 Section 2, we describe the SDSS data and the Cut \& Enhance cluster
 catalog. In Section 3, we analyze the late type fraction,
 both spectrally and morphologically. In Section 4, we discuss
 the possible caveats and underlying physical processes in the evolution of galaxies.  In
 Section 5, we summarize our work and findings.
 The cosmological parameters adopted throughout this paper are $H_0$=75 km
 s$^{-1}$ Mpc$^{-1}$, and ($\Omega_m$,$\Omega_{\Lambda}$,$\Omega_k$)=(0.3,0.7,0.0).

\section{Data}\label{data}

  The galaxy catalog used here is taken from the Sloan Digital Sky Survey (SDSS)
 Early Data Release (see Fukugita et al. 1996, Gunn et al. 1998,  Lupton
 et al. 1999, 2001, 2002, York et al. 2000, Hogg et al. 2001, Pier et al. 2002, Stoughton et al. 2002 and Smith et al. 2002 for more detail of the SDSS data). 
 We use equatorial scan data, a contiguous area of 250 deg$^2$
 (145.1$<$RA$<$236.0, -1.25$<$DEC$<$+1.25) and 150 deg$^2$
 (350.5$<$RA$<$56.51, -1.25$<$DEC$<$+1.25).
 The SDSS imaging survey observed the region to depths of 22.3, 23.3, 23.1, 22.3
 and 20.8 in the $u,g,r,i$ and $z$ filters, respectively. (See
 Fukugita et al. 1996 for the SDSS filter system, Hogg et al. 2002
 and Smith et al. 2002 for its calibration). 
 Since the SDSS photometric system is not yet finalized, we refer to the
 SDSS photometry presented here as $u^*,g^*,r^*,i^*$ and $z^*$. We
 correct the data for galactic extinction determined from the maps given
 by Schlegel, Finkbeiner \& Davis (1998).  
 We include galaxies to $r^*$=21.5 (petrosian magnitude), which is the
 star/galaxy separation limit (studied in detail by Scranton et
 al. 2002) in the SDSS data. 

 The galaxy cluster catalog used here is a subset of the SDSS Cut \& Enhance
 galaxy cluster catalog (Goto et al. 2002a). There are 4638 clusters in
 the equatorial region (See Kim et al. 2002, Annis et al. 2002, Bahcall
 et al. 2002 and Miller et al. 2002 for other works on the SDSS
 galaxy clusters). Besides the uniformity of the catalog with its well
 defined selection function, the catalog has very good photometric redshifts,
 $\delta z$=0.015 at $z<$0.3, which enables us to use a large sample
 of clusters (Goto et al. 2002a; see also Gal et al. 2000 and Annis et
 al. 2002 for photometric redshift methods for clusters). We use
 clusters in the redshift range of 0.02$\leq$z$\leq$0.3 
 and galaxies brighter than $M_r=-19.44$, which corresponds to $r^*$=21.5
 at z=0.3. 
 Since several authors in previous work claimed that
  biases in sample cluster selection can mimic the evolutionary effect, it
 is important to control cluster richness well. We use clusters with
 more than 25 member galaxies between $M_{r^*}=-24.0$ and  $M_r=-19.44$
 within 0.7 Mpc 
 after fore/background subtraction, as explained in the next
 Section. The large areal coverage of the SDSS data 
 enables us to subtract the fore/background counts reliably. We can thus
  control the richness of the sample clusters well. The criteria
 leave 514 clusters in the region.  

 We stress the importance of the uniformity of the cluster
 catalog. Although the Abell cluster catalog (Abell 1958, Abell, Corwin
 and Olowin 1989) has been used in many studies, it was constructed by eye,
 and is sensitive to projection effects.
 When comparing clusters at different redshifts, it is particularly important
 to ensure that the data quality and cluster selection techniques are
 comparable, to avoid the introduction of potential selection biases.
%
 The SDSS Cut \& Enhance cluster catalog used here is constructed using
 only the 
 data taken with the SDSS telescope. Also, clusters
 are detected using a single algorithm (CE) throughout the
 entire redshift range (0.02$\leq$z$\leq$0.3). Combined with the well controlled
 richness criteria, our cluster sample is not only one of the biggest
 but also one of the most statistically uniform cluster catalogs. 
  To study colors of galaxies in clusters, it is also important to use a
 cluster catalog created without targeting the red sequence of color
 magnitude relation of  cluster galaxies. For example, Gladders \&
 Yee(2000) and Annis et 
 al. (2002) use a color filter targeting the red sequence of clusters
 and find galaxy clusters efficiently without suffering from 
 projection effects. These techniques, however, can potentially have
 biases with regards to the colors of detected galaxy clusters, since
 clusters with a strong red sequence 
 is more easily detected.  They may not, therefore, be ideally
 suited to a Butcher-Oemler type of analysis.   In contrast, the SDSS Cut \& Enhance
 cluster catalog does not pick red galaxies selectively, and is therefore more
 suitable for this study. (Note that the the Cut \& Enhance method does
 use generous color cuts. Although the color cut is designed to be wide
 enough to include blue galaxies in clusters, 
 therefore it is not completely free from color originated biases.)  In
 previous work, 
 clusters have often been detected using data from only one or two
 color bands.  This can introduce a bias, since
 higher redshift clusters are redder and fainter than lower
 redshift clusters. 
 The SDSS Cut \& Enhance cluster catalog detects clusters using
 four bands of the SDSS data ($g,r,i$ and $z$), which minimizes the bias
 against redshift. 


\section{Analysis and Results}\label{method}

\subsection{Fore/Background subtraction}\label{bg}

 We compute fractions of late-type
 galaxies in four different ways. 
 First, we describe the statistical fore/background
 subtraction method we use.  
 In counting galaxies, all galaxies are assumed to be at the cluster
 redshift to calculate absolute magnitudes. Then galaxies only between
 $M_{r^*}=-24.0$ and $M_{r^*}=-19.44$ are used in the analysis. 
 We count the number of late-type/total galaxies within 0.7 Mpc from each
 cluster. 
 Valotto et al. (2001) claimed that the global background
  correction can not correct background contamination appropriately.
   Following the claim, we use a local background correction.  The
  number of late-type/total galaxies in fore/background is estimated
  using an annular area around each cluster with an inner 
 radius of 2.1 Mpc and an outer radius of 2.21 Mpc in
 the same absolute magnitude range.    
  The annulus fore/background subtraction enables us to
 estimate the fore/background locally, minimizing variations in galaxy
 number counts due to the large scale structure. 
  When an outer annulus touches the boundary of the region, a
 fore/background count is globally subtracted using galaxy number counts
 in the entire 400 deg$^2$ region by adjusting it to the angular area each
 cluster subtends. 
 This fore/background subtraction is used in the analyses described in
   sections 3.2-3.5. 
 The fraction of blue/late-type galaxies and its error are computed
 according to the following equations.
  \begin{equation} 
    f_{late}=\frac{N_{c+f}^{late}-N_{f}^{late}}{N_{c+f}^{all}-N_{f}^{all}},\label{frac}
 \end{equation}
  \begin{equation} 
    \delta f_{late}=   f_{late}\times \sqrt{\frac{ N_{c+f}^{late}+N_{f}^{late}}{(  N_{c+f}^{late}-N_{f}^{late}  )^2}  +    \frac{N_{c+f}^{all}+N_{f}^{all}}{(N_{c+f}^{all}-N_{f}^{all})^2}   - \frac{2(\sqrt{N_f^{late}\times N_f^{all}}+ \sqrt{N_{c+f}^{late} \times N_{c+f}^{all}})}{(N_{c+f}^{late}-N_f^{late})(N_{c+f}^{all}-N_f^{all})} }
\label{frac_err}
 \end{equation}
 \noindent where $N_{c+f}^{late}$ and $N_{f}^{late}$ represent numbers
 of blue/late-type galaxies in a cluster region and a field region,
 respectively. $N_{c+f}^{all}$ and $N_{f}^{all}$ represent numbers
 of all galaxies in a cluster region and a field region, respectively.
 The equation (\ref{frac_err}) assumes that $N^{late}$ and $N^{all}$ are not
 independent. We explain the derivation of  equation (\ref{frac_err}) in
 Appendix \ref{error}.

\subsection{Blue Fraction}\label{f_b}
  
 The blue fraction of galaxy clusters ($f_b$) is measured as the fraction
 of galaxies bluer in $g-r$ rest frame color than the color of the ridge
 line of the cluster by 
 0.2 mag. This color criterion is equivalent to
 Butcher \& Oemler's (1984) 0.2 mag in $B-V$ and Margoniner et al.'s (2000,
 2001) 0.2 mag in $g-r$.  
 The color of the ridge line is measured from the color-magnitude
 diagram using the same color-magnitude box used in measuring
 photometric redshift (Goto et al. 2002a). 
 The colors of the ridge lines are confirmed
 to agree with empirical color of elliptical galaxies  observed in the
 SDSS at the same redshift with less than 0.05 difference in $g-r$ color
 (Eisenstein, private communication). 
 We also use Fukugita et
 al.'s (1995) model of an elliptical galaxy and a galaxy bluer than it by
 0.2 mag in  $g-r$. By redshifting these two galaxies, we measure
 $\delta (g-r)$ in the observed frame, which corresponds to the
 restframe  $\delta (g-r)$=0.2 . In calculating $f_b$, we count 
 galaxies within 0.7 Mpc from the center of each cluster, which is the
 same radius as Margoniner et al. (2000, 
 2001), and corresponds to the average radius of Butcher \& Oemler
 (1984). (We explore possible caveats in using fixed radius in Appendix.)
 Galaxies between $M_{r^*}=-24.0$ and $M_{r^*}=-19.44$ are counted, which
 corresponds to $r^*$=21.5 at $z$=0.3 for an average $k$-correction of all types of
 galaxies (Fukugita et al. 1995). Compared to the field luminosity function
 of Blanton et al. (2001), this includes galaxies as faint as $M^*_r$+1.36.  
 Fore/background galaxies are statistically subtracted in the way described in Section \ref{bg}.
  The lower left panel of Figure \ref{fig:bo} shows $f_b$ as a function of  redshift. 
 The error in $f_b$ is estimated using equation (\ref{frac_err}) and
 the median values of the errors in $f_b$ and z are shown in the upper
  left corner of the plot.  
 Dashed line shows the weighted least-squares fit to the data.
 Solid lines and stars show the median values of the data. 
 The scatter is considerable, but both of the lines show a clear increase of $f_b$
 toward higher redshift. The  Spearman's correlation coefficient is
 0.238 with significance of more than 99.99\% as shown in Table
  \ref{tab:correlation}. The correlation is weak, but
 of high significance.  The lower left panel of Figure \ref{fig:kstest}
  further clarifies the evolution effect. A dashed and a solid line show 
  normalized distributions of $f_b$ for clusters with z$\leq$0.15 and
  0.15$<$z$\leq$0.3, respectively. The two distributions are significantly
  different at the 98\% level, as determined by a Kolomogorov-Smirnov test.
 The slope shown with the dashed line
 rises up to $f_b\sim$=0.2 at $z$=0.3 (look back time of $\sim$3.5 Gyr),
 which is consistent with previous
 work such as Butcher \& Oemler (1978,1984), Rakos \& Schombert (1995),
 Margoniner et al. (2000, 2001), within the scatter. 
 We conclude that the Butcher-Oemler effect is seen in 
 the SDSS Cut \& Enhance galaxy cluster catalog.

   We caution readers on the systematic uncertainties in measuring $f_b$.
  Marzke et al.(1994,1997,1998), Lin et al. (1999) and Blanton et
  al. (2001)  showed that luminosity functions of galaxy clusters depend
  significantly on galaxy type, in such a way that
  the bright end of the  cluster luminosity function is dominated by
  redder galaxies and the  faint end is dominated by bluer galaxies. 
  Boyce et al.(2001) and Goto et al. (2002b) showed that a similar
  tendency exists for cluster luminosity functions. 
  This difference in luminosity functions leads to a different blue fraction depending
  on  the absolute magnitude range used. Furthermore, if the radial
  distributions of blue and red galaxies are different ($e.g.$ Kodama et
  al. 2001), the $f_b$ measurement
  depends heavily on the radius.  When comparing with previous work, therefore,
  it is important to take account of the exact method used to calculate $f_b$.
  We discuss the
  uncertainty in measuring blue fractions further in Section \ref{discussion}.

\subsection{Late Type Fraction Using $u-r<$2.2}\label{u-r}

 Recently Shimasaku et al. (2001) and Strateva et al. (2001) showed that
 the SDSS $u-r$ color correlates well with galaxy morphologies.
 In this section we use $u-r$
 color to separate early($u-r\geq$2.2) and late ($u-r<$2.2) type galaxies
 as proposed by Strateva et al. (2001). Note that although $u-r$ color
 is claimed to correlate well with galaxy types, it is still a color
 classifier and thus different from the morphological parameters
 we investigate in the following two sections.
 The methodology used to measure late type fraction is similar to the one we use
 to measure $f_b$. We regard every galaxy with $u-r<$2.2 as a late type
 galaxy. We define $f_{u-r}$ as the ratio of the number of late type
 galaxies to the total number of galaxies within 0.7 Mpc.
 Fore/background subtraction is performed in a way described in Section \ref{bg}.   
 The upper left panel of figure \ref{fig:bo} shows $f_{u-r}$
 as a function of redshift. The error in $f_{u-r}$ is calculated using equation
 (\ref{frac_err}) and the median  values of the errors in $f_{u-r}$ and z
 are shown in the upper left corner
 of the
 panel. The dashed line shows the least square fit to the data.
   The solid lines and stars show the median values of the data. 
 As in the case of $f_b$, the scatter is considerable, but the weak
 increase of the late type galaxies is seen. The  Spearman's correlation coefficient
 is 0.234 and is inconsistent with zero at greater than 99.99\% confidence
 level (Table \ref{tab:correlation}). Again, weak but significant
 correlation is found. The upper left panel of Figure \ref{fig:kstest}
 shows distributions of $f_{u-r}$ for z$\leq$0.15 clusters and
 0.15$<$z$\leq$0.3 clusters with a dashed and solid line, respectively. A Kolomogorov-Smirnov
 test shows that the distributions are different with more than a 99\%
 significance. 
 In addition to the increase in $f_b$ shown in the last Section,
 the increase in $f_{u-r}$ provides further evidence of color
 evolution of cluster galaxies.  Furthermore, since $u-r$ color of galaxies is
 sensitive to a galaxy's morphology as shown in Figure 6 of Strateva et
 al (2001), it suggests possible evolution of morphological types of galaxies
 as well.
 We investigate the morphological evolution of galaxies in clusters in the next subsection.

\subsection{Late Type Fraction Using Profile fitting} \label{exp}

\begin{figure}[h]
\begin{center}
\includegraphics[scale=0.7]{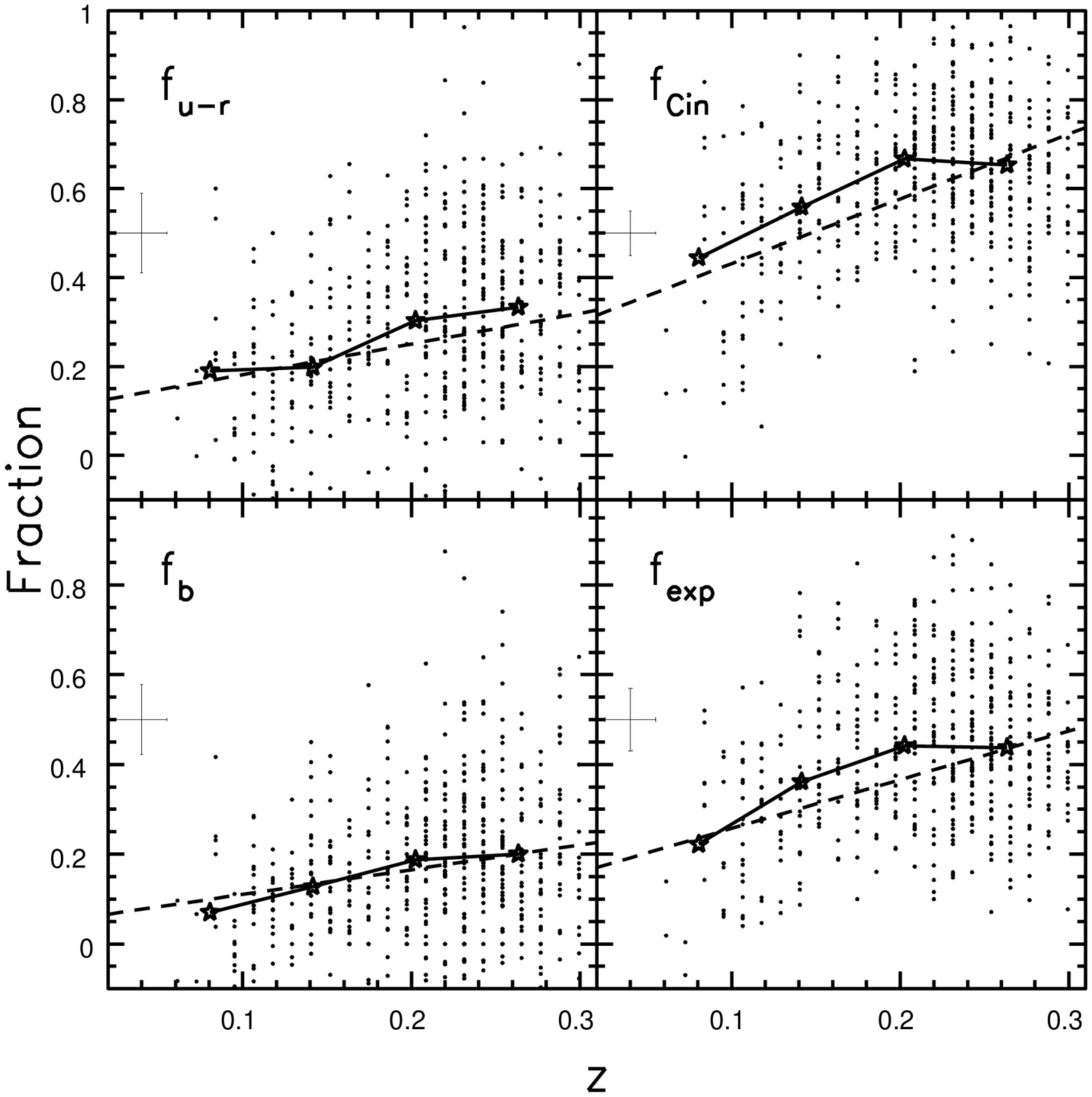}
\end{center}
\caption{\label{fig:bo}
 Photometric and morphological Butcher-Oemler effect from the 514 SDSS Cut \&
 Enhance clusters. 
$f_b$, $f_{Cin}$, $f_{exp}$ and $f_{u-r}$ are plotted against redshift. 
The dashed lines show the weighted least-squares fit to the data. The stars
 and solid lines show the median values. The median values of errors
 are shown in the upper left corners of each panel. The Spearman's correlation coefficients are shown in Table \ref{tab:correlation}.
}
\end{figure}

\begin{figure}[h]
\begin{center}
\includegraphics[scale=0.7]{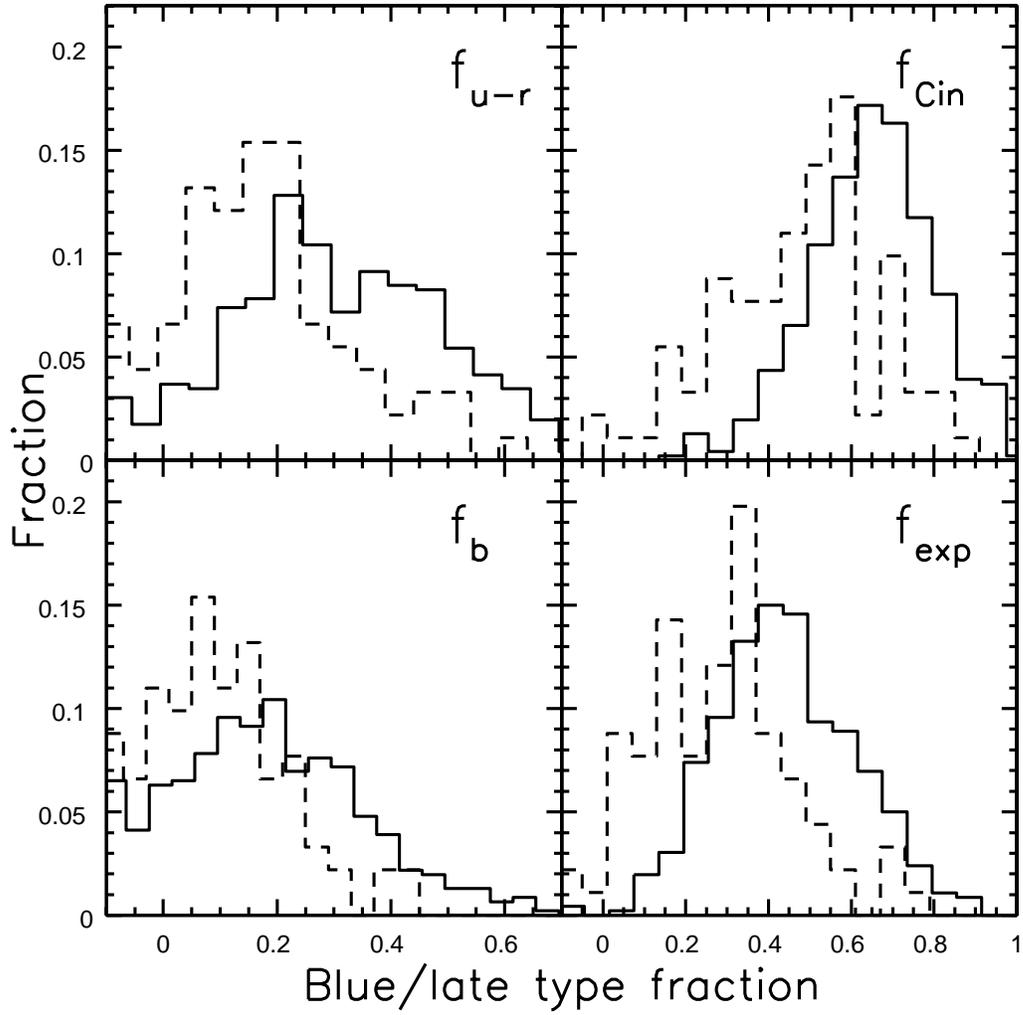}
\end{center}
\caption{\label{fig:kstest}
 Normalized distributions of late type fractions ($f_b$, $f_{Cin}$, $f_{exp}$ and $f_{u-r}$).
 Dashed lines show distributions of lower redshift clusters
 (z$\leq$0.15) and solid lines show ones of higher redshift clusters
 (0.15$<$z$\leq$0.3).
 The results of Kolomogorov-Smirnov tests are shown in Table \ref{tab:kstest}. In all
 cases, Kolomogorov-Smirnov tests show the two distributions are significantly different. 
}
\end{figure}

  One of the purposes of this paper is to determine if there is a
  morphological change of galaxies in clusters as a function of
  redshift. 
  The SDSS photometric pipeline (PHOTO; Lupton et al. 2002) fits a de Vaucouleur
 profile and an exponential profile to every object detected in the SDSS imaging
 data and returns the likelihood of the fit. By comparing the likelihoods
 of having an exponential profile against that of a de Vaucouleur profile, we can classify
 galaxies into late and early types. In this section, we
 regard every galaxy that has an exponential likelihood higher than a de Vaucouleur
 likelihood in $r$ band as a late type galaxy. A galaxy with higher
 de Vaucouleur likelihood in $r$ band is regarded as an early type
  galaxy. 
 We define $f_{exp}$ in the same way as in previous subsections. $i.e.$
 $f_{exp}$ is the ratio of the number of late type galaxies to
 the total number of galaxies within 0.7 Mpc. Fore/background counts are
  corrected using the method described in Section \ref{bg}. 
 The resulting  $f_{exp}$ is plotted in the lower right panel of Figure
 \ref{fig:bo}. 
  The error in $f_{exp}$ is estimated using equation (\ref{frac_err}) and
 the median values of the errors in  $f_{exp}$ and z are shown in the upper
  left corner of the plot.  
 The dashed line shows the weighted least-squares fit. 
 The solid lines and stars show the median values of the data. 
 The scatter is considerable, but we see the increase of $f_{exp}$ toward the higher redshift. The
  Spearman's correlation coefficient is 0.194, which is inconsistent with zero at
 more than a 99.99\% confidence level (Table \ref{tab:correlation}).
 The upper right panel of Figure \ref{fig:kstest} shows the
 distributions of clusters with z$\leq$0.15 and
 with 0.15$<$z$\leq$0.3 with a dashed and solid line, respectively. 
 The two distributions show a difference of more than 99\% significance
 in a Kolomogorov-Smirnov test. 
 We emphasize that the galaxy
 classification used here is purely morphological --- independent
 of colors of galaxies. The fact that we still see the increase of the
 late type galaxies toward higher redshift suggests that these
 Butcher-Oemler type blue galaxies also change their morphological
 appearance as well as their colors. We also point out that the slope of the
 change is similar to the panel in the lower left of Figure
 \ref{fig:bo}, which is  $\sim$30\% between z=0.02 and z=0.3. 
 We note that there is a potential bias associated with the use of $r$
 band profile fitting throughout 
 the redshift range, since the $r$ band wavelength range at $z$=0.3 is almost that of $g$
 band at restframe. 
 We investigate this effect in Section \ref{discussion}, and conclude that it is small. 
 Like the blue fraction, the morphological late-type fraction is also sensitive to the
 magnitude range considered.
  Binggeli et al. (1988), Loveday et al. (1992), Yagi et al. (2002a,b) and Goto et al. (2002b) reported
 luminosity functions of elliptical galaxies have brighter
 characteristic magnitudes and flatter faint end tails compared to those
 of spiral galaxies in both field and cluster regions. Careful attention should be paid to 
 the magnitude range used in an analysis when fractions of spiral galaxies are compared. 
  We discuss the uncertainty further in Section \ref{discussion}.

\subsection{Late Type Fraction Using Concentration Parameter} \label{Cin}

   As another morphological galaxy classification method, we use the inverse of
   the concentration parameter ($C_{in}$) advocated by Shimasaku et
   al. (2001) and Strateva et al. (2001). We define $C_{in}$ as the
   ratio of Petrosian 50\% radius to Petrosian 90\% radius in $r$
   band. They are the radii which
   contain 50\% and 90\% of Petrosian flux, respectively. (See Stoughton
   et al. 2002 for more details of Petrosian parameters). Since $C_{in}$
   is the inverse of a conventional concentration parameter, spiral
   galaxies have a higher value of $C_{in}$. Following Strateva et al. (2001), we
   use $C_{in}$=0.4 to divide galaxies into early and late type
    galaxies. Readers are referred to Morgan (1958,1959), Doi, Fukugita \& Okamura (1993) and
   Abraham et al. (1994, 1996) for previous usage of concentration of light
   as a classification parameter.    
   $f_{Cin}$ is defined as the ratio of the number of galaxies
   with $C_{in}>$0.4 to the total number of galaxies within 0.7 Mpc as
   in the previous subsections. Note that our early type galaxies with
   $C_{in}<$0.4 include S0 galaxies in addition to elliptical galaxies
   since discerning elliptical and S0 galaxies is very difficult with
   the SDSS data, in which the seeing is typically 1.5''. (See Shimasaku et
   al. 2001 and Strateva et al. 2001 for the correlation of $C_{in}$
   with an eye classified morphology).
   Fore/background
   number counts are corrected as described in Section \ref{bg}.
   The absolute magnitude range used is  
   $-24<M_r<-19.44$. The upper right panel of Figure \ref{fig:bo} shows $f_{Cin}$  as a function
   of redshift. Since the classification using $C_{in}$=0.4 leans
   toward late type galaxies, the overall fraction is higher than the
   other panels in the figure. The increase of late type fraction,
   however, is clearly seen. The dashed line shows the weighted least-squares 
   fit.  The solid lines and stars show the median values of the data. 
  The error in $f_{Cin}$ is estimated using equation (\ref{frac_err}) and
 the median values of the errors in  $f_{Cin}$ and z are shown in the upper
  left corner of the plot.  
 The  Spearman's correlation coefficient is
 0.223 with significance of more than 99.99\% as shown in Table
   \ref{tab:correlation}. The upper right panel of Figure
   \ref{fig:kstest} further clarifies the evolution effect. The
   distribution of z$\leq$0.15 clusters in a dashed line and the
   distribution of 0.15$<$z$\leq$0.3 clusters in a solid line show a
   difference with more than a 99\% significance level.   
   We stress that the galaxy classification
   based on this concentration parameter is purely a morphological one.
  In this morphological classification, we still see the increase of the late type fraction just
   like the increase of $f_b$ --- as if observing the morphological equivalence of the
   Butcher-Oemler effect.
   The increase in $f_{Cin}$ combined with the increase in $f_{exp}$ 
   provides rather firm evidence of morphological change in the
   Butcher-Oemler type
   galaxies. Possible caveats in the usage of $C_{in}$ and comparisons
   with previous works are discussed in Section \ref{discussion}.

\subsection{On the Origin of the Scatter} \label{scatter}

\begin{figure}[h]
\begin{center}
\includegraphics[scale=0.7]{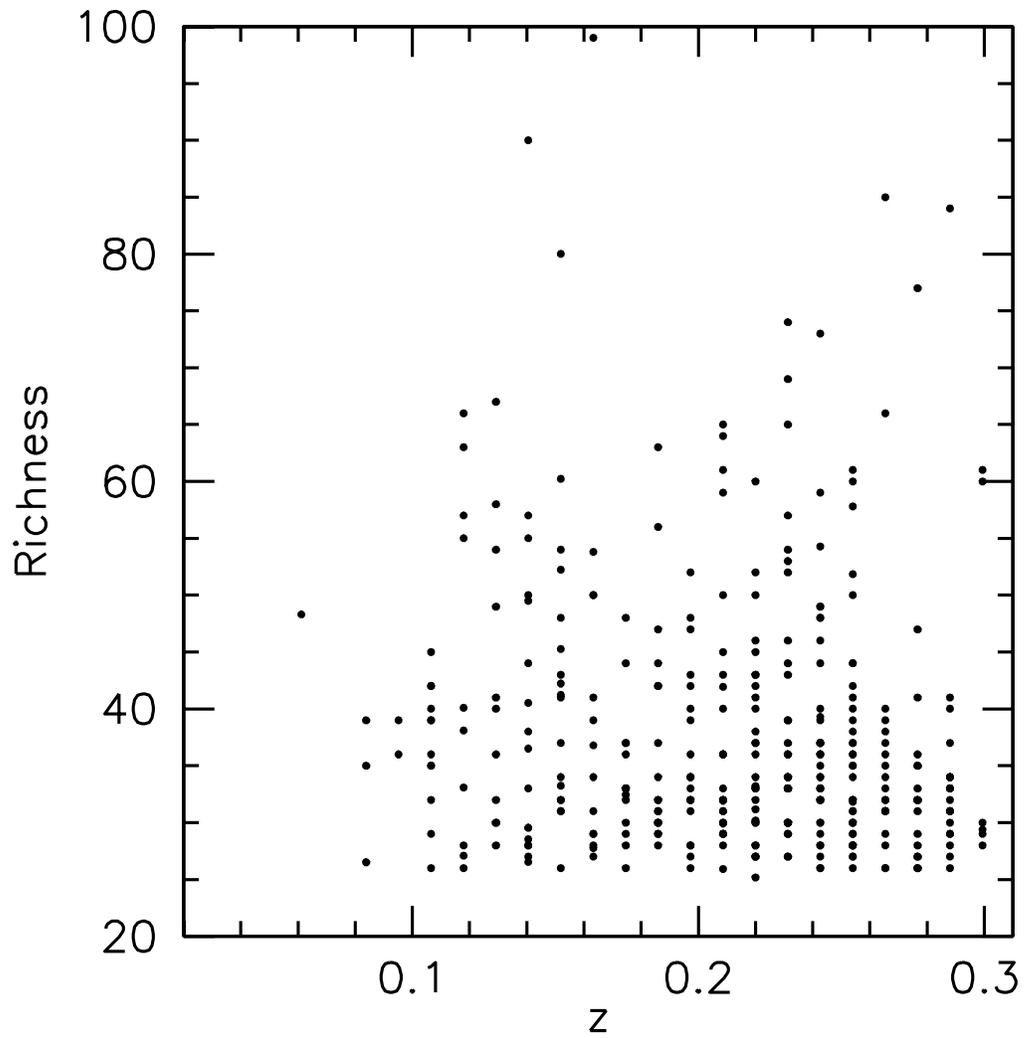}
\end{center}
\caption{
\label{fig:z_rich}
 Richness distribution as a function of redshift. Richnesses are measure
 as the number of galaxies brighter than $M_r=-19.44$ within 0.7 Mpc after
 fore/background subtraction. 
}
\end{figure}

 In the last four sections, we observed the increase of late type
 fractions toward higher redshift in all the four cases. At the same time, we see
 a significant amount of scatter around the late type fraction
 v.s. redshift relations. Although the errors on these measurements are
 also large, this scatter might suggest that there might be one or more physical
 properties  which determine the amount of late-type galaxies in
 clusters. Table \ref{tab:error_comparison} compares median error sizes
 of $f_b,f_{u-r},f_{exp}$ and $f_{Cin}$ with scatters around the
 best-fit lines (dotted lines in Figure \ref{fig:bo}). 
 In fact, in all cases, real scatters are larger than the statistical errors.
 In the literature, several 
 correlations are proposed  such as with
 X-ray shapes of clusters (Wang et al. 2001, Metevier et al. 2000), and 
 cluster richness (Margoniner et al. 2000). 
 Our cluster richnesses are plotted against redshift in Figure \ref{fig:z_rich}.
 Richnesses are measured as numbers of galaxies between $M_r=-24.0$ and
 $M_r=-19.44$ 
 within 0.7 Mpc after fore/background subtraction, as explained in
 Section \ref{bg}. In figure \ref{fig:z_rich}, this richness has no
 apparent bias with redshift. 
 In Figure \ref{fig:okamura_rich}, the difference of the late type
 fraction from the best-fit line is plotted against cluster richness.
 The circles and solid lines show median values. 
 In all the panels, there is a clear tendency of richer clusters 
 having a lower fraction of late type galaxies. This tendency is in
 agreement with Margoniner et al. (2000) who found richer clusters
 had lower blue fractions. We further discuss the richness
 dependence of the late type fraction in Section \ref{discussion} and Appendix.
  As an alternative parameter to X-ray shape, we plot the difference from
 the best-fit line against cluster elongation in Figure
 \ref{fig:okamura_elong}. The elongation parameter is taken from Goto
 et al. (2002a), which measured the ratio of major and minor axes
 in  their enhanced map to find clusters. Circles and solid lines show
 median values. No obvious trend is seen here. Our result
 seems to agree with Smail et al.'s (1997) caution that a correlation between $f_b$
 and cluster ellipticity found by Wang et al. (1997) could be due to a small
 and diverse sample.
 However, since distribution of
 galaxy positions might not represent cluster ellipticities well, we do not
 conclude  that there is no dependence on cluster ellipticities. The
 dependence should be pursued further in the future, ideally using X-ray
 profile shape with a large sample of clusters.

\section{Discussion}\label{discussion}

\subsection{Morphological k-correction}

 In the upper right panel of Figure \ref{fig:bo}, we use $C_{in}$
 (inverse of concentration index) in the $r$ band to classify galaxies throughout our redshift
 range (0.02$\leq$z$\leq$0.3). This could potentially cause redshift dependent
 biases in our calculation of $C_{in}$. Since the universe is expanding,
 by analyzing the observed $r$ band data, we are analyzing bluer
 restframe wavelengths in the higher redshift galaxies.
 In fact, the $r$ band at $z$=0.3 is almost $g$ band in the
 restframe.  Various authors have pointed out that galaxy
 morphology significantly changes according to the wavelength used
 ($e.g.$ Abraham et al. 2001). To estimate how large this bias is, we
 plot the normalized distributions of $C_{in}$ in $g$ and $r$ bands  in
 Figure \ref{fig:cin_g_r} using
 the galaxies with 0.02$\leq$z$\leq$0.03 in the SDSS spectroscopic data (1336
 galaxies in total; See Eisenstein et al. 2001, Strauss et al. 2002 and
 Blanton et al. 2002 for the SDSS spectroscopic data). In this redshift range, the color shift due to the
 expansion of the universe is small. We use this to study the
 dependence of the $C_{in}$ parameter on the restframe wavelength. At
 $z$=0.3, $r$ band corresponds to restframe $g$ band. The solid
 and dashed lines show the distribution for $g$ and $r$ bands,
 respectively. The two distributions are not exactly the same, but the
 difference between the two distributions is small. We summarize the
 statistics in Table \ref{tab:cin_g_r}. 
 There are 802/1336 galaxies with $C_{in}>$0.4 in $g$ band, and 787/1336
 galaxies have $C_{in}>0.4$ in $r$ band. The difference is 15/1336 galaxies, which is 1.1\% of the
 sample.  In section  \ref{Cin}, the change in $f_{Cin}$ is $\sim$30\%.
 The effect of the morphological k-correction is therefore much smaller.  
 We also point out that this analysis assumes the largest difference in
 redshift (0.02$\leq$z$\leq$0.3),
 therefore it gives the upper limit of the bias. 
 Since the majority of our clusters are at 
 $z\sim$0.2, the wavelength difference between the observed and
 restframe bands is typically much smaller.
 We conclude that the effect of the morphological k-correction
 is much smaller than the change in $f_{Cin}$ we observed in Section \ref{Cin}.

  In Section \ref{exp}, we use the $r$ band fit for all galaxies in our
 sample. The same redshift effect could potentially bring bias to our
 analysis. In Table \ref{tab:exp_g_r}, we limit our galaxies to
 0.02$\leq$z$\leq$0.03 and count the fraction of late type galaxies in
 the $g$ and $r$ bands corresponding to observed frame $r$ band at z=0.0 and z=0.3. 
 We list the number of galaxies with
 exponential likelihood higher than de Vaucouleur likelihood in column 1, the
 total number of galaxies in column 2, and the ratio of columns 1 to 2 in column
 3. As shown in the 3rd row, the difference in the fraction of late type
 galaxies between $g$ band data and $r$ band data is only 2.5\%, which
 is much smaller than the $f_{exp}$ change we see in the upper right panel of
 Figure \ref{fig:bo} ($\sim$30\%). 
 We conclude that the change of
 $f_{exp}$ and $f_{Cin}$ is not caused by the small redshift bias in using $r$ band
 data throughout the redshift range. 
%


\subsection{Seeing Dependence}

  Another possible source of bias in measuring $f_{exp}$ and
 $f_{Cin}$ is the dependence on the seeing, relative to the size of the galaxies. At
 higher redshift, the size of a galaxy is smaller and a seeing convolution
 could be more problematic.  Especially for the concentration
 parameter ($C_{in}$), galaxy light becomes less concentrated when
 the seeing size is comparable to the galaxy size, and thus,  the effect could
 cause a bias towards higher $C_{in}$ values.  
 To check this, we plot $f_{Cin}$ against the
 point-spread function (PSF) size in the $r$ band for two redshift limited samples in Figure
 \ref{fig:seeing_cin}.  Open squares and solid lines show
 the distribution and medians of low z clusters (z$\leq$0.15). Filled
 triangles and dashed lines show the distributions and medians of high z
 clusters (0.24$<$z$\leq$0.3). 
 For the median measurements, bins are chosen so that equal
 numbers of galaxies are included in each bin.  1 $\sigma$ errors are shown as vertical bars.
 As expected, lower
 redshift clusters show almost negligible dependence on seeing
 size. Higher redshift clusters show about a 5\% increase in $f_{Cin}$ between the best
 and worst seeing size. The evolution effect we see in the upper right
 panel of Figure \ref{fig:bo} is more than 20\%. Furthermore, 
 as the distribution of seeing in Figure \ref{fig:seeing_hist} shows,
 87\% of our sample galaxies have seeing better than 2.0''. 
 Therefore we conclude that varying seeing causes
 a small bias which is significantly weaker than
 the evolution we find in Section \ref{method}.   
 The effect of varying seeing is less significant for the $f_{exp}$
 parameter. In Figure \ref{fig:seeing_exp}, we plot $f_{exp}$ against
 seeing size for two redshift samples with redshift ranges and symbols
 as in Figure \ref{fig:seeing_cin}.  1 $\sigma$ errors, shown as
 vertical bars, are dominant.  There is no significant correlation of
 $f_{exp}$ with seeing size.

\subsection{Radius, Fore/background Subtraction and Cluster Centroids}
 
  Throughout the analyses in Section \ref{method}, we use 
 a 2.1-2.21 Mpc annular region for fore/background
 subtraction. In return for
 taking cosmic variance into account, annular (local) background subtraction  
 has larger errors than global background subtraction due to its smaller
 angular area coverage. However the difference is not so large. In
 case of blue galaxy counts ($f_b$) in the background, the
 median Poisson (1 $\sigma$) uncertainty for global background is 12.2\%, whereas 1
 $\sigma$ variation of local background is 12.6\%. This increases
 the errors, but only by 0.4 points. The actual effect to the late-type
 fraction is plotted in Figure \ref{fig:bg_test}.  Solid lines show
 distributions for our default choice of 0.7 Mpc radii 
 and 2.1-2.21 Mpc annuli. Dashed lines show distributions for global
 fore/background subtraction, where fore/background subtraction is
 performed using global number counts of galaxies for all the clusters
 in the sample. A Kolomogorov-Smirnov test between two samples does not show any
 significant difference.

  For cluster radius, we use 0.7 Mpc, since we do not have
 information about the virial radii of each system.
 It is, however, ideal to use virial radii since, for example, in a standard cold dark
 matter cosmology, virial radii
 at a fixed mass scales as $\propto$ (1+z)$^{-1}$. 
  Another possible cause of uncertainty is the accuracy in deciding cluster
 centers. 
 In this work, a center position of each cluster is taken from
 Goto et al. (2002a), and is estimated from the position of the peak in their
 enhanced density map. Although, from Monte-Carlo simulations, cluster centroids are expected
 to be determined with an accuracy better than $\sim$40 arcsec,
 the offsets have a possibility to introduce a bias in our analyses.
 We test 
 different choices of these parameters in Figure \ref{fig:bg_test}. 
 Dotted lines show distributions where radii change as
 0.7 $\times$ (1+z)$^{-1}$ Mpc assuming a standard cold dark matter
 cosmology.  Long dashed lines show distributions when the position of
 brightest cluster galaxy (within 0.7 Mpc and  $Mr<$-24.0) is used as a cluster center. 
 Kolomogorov-Smirnov tests show no significant difference in any of the above cases.
 In all cases, the probability that the distributions are different is
 less than 26\%.  Our results in Section \ref{method} are thus not
 particularly sensitive to our choice of annuli, radii  or cluster centers.
 We further pursue the effect of radius dependence of blue/late type fractions in
 Appendix, and show that it does not change our main results.

\subsection{Comparison with Late-type Fraction from Spectroscopy}

 To further test our late-type fraction measurement, we compare
 the late-type fraction obtained from the SDSS spectroscopic data with
 that obtained from the SDSS imaging data in figure
 \ref{fig:comparison_with_spectroscopy}. Since the SDSS spectroscopic
 data are limited to $r^*<$17.77 (Strauss et al. 2002), the comparison
 can be done only for clusters with z$<$0.06. In the literature, three
 clusters are found to satisfy these criteria in the region used in this
 study. The clusters include ABELL 295, RXC
 J0114.9+0024, and  ABELL 957. For these clusters, late-type fractions
 are measured in the same way as in section \ref{method}. Late-type
 fractions from spectroscopy are measured using galaxies within 0.7 Mpc
 and $\delta$z=$\pm$0.005 from the redshift of each cluster. Note that there is no
 fore/background correction for spectroscopic late-type fraction. In figure
 \ref{fig:comparison_with_spectroscopy}, all points agree with each
 other within the error. The good agreement suggests that our fore/background
 subtraction technique described in section \ref{bg} works
 properly. It would be ideal to perform the same test for high redshift
 clusters as well. However, the SDSS spectroscopic data are not deep
 enough to perform the test for higher redshift clusters.

\subsection{The Butcher-Oemler Effect}

  The Butcher-Oemler effect--- an increase in the ratio of blue galaxies in
  clusters as a function of redshift--- is strong evidence of direct evolution of
  the stellar populations in galaxies; it has been studied by numerous
  authors in the past.
 In this section,
 we compare our results with previous work.
 Since different authors use  
 different cluster samples, color bands, cosmology, absolute
 magnitude ranges and methods of fore/background subtraction,
 which could affect the comparison,
 we emphasize the differences in analysis by each author. Note that one
 important difference is that some previous work used
 a sample of quite rich clusters.  e.g. clusters with more than 100 members
 in magnitude and radius ranges comparable to those adopted in this
 study. Poorest systems in our 
 sample have only 25 member galaxies after fore/background subtraction. 
 Thus, difference in cluster samples could cause a difference in results.

  Butcher \& Oemler (1978, 1984) studied 33 clusters between z=0.003 and
 z=0.54. They used galaxies brighter than $M_V$=-20 ($h$=0.5 and $q_0$=0.1)  within the circular area
 containing the inner 30\% of the total cluster population. They found
 $f_b$ increases with redshift for z $\geq$ 0.1. Their $f_b$ at z=0.3 is
 $\sim$0.15, which is slightly lower than our value. 
 Considering the large scatter in both their and our samples, we do not claim that our results
 are inconsistent with their value. Note that Andreon \& Ettori (1999)
 found a trend of increasing X-ray luminosity with increasing redshift
 in the sample clusters of Butcher \& Oemler (1984).    
 
 Rakos \& Schombert (1995) studied 17 clusters using Stromgren $uvby$
 filters. Due to the usage of the narrow band filters redshifted to the
 cluster distance, their study did need to use model-dependent k-corrections.
 However, their high-redshift cluster sample is drawn from that
 of Gunn, Hoessel \& Oke (1986) which is based on IIIa-J and IIIa-F
 photographic plates.  At $z>0.5$, these plates measure the rest-frame
 ultraviolet to blue region of the spectrum.  Thus the cluster catalog will
 be biased toward clusters rich in blue galaxies at high redshift.
 Rakos \& Schombert found $f_b\sim$0.25 at z=0.3, which is
 slightly higher than the estimation of Butcher \& Oemler (1984) but
 in agreement with our results. 

 Margoniner et al. (2000) studied 44 Abell clusters between z=0.03 and z=0.38. They used galaxies
 between $M_r=-21.91$ and $M_r=-17.91$ ($h=0.75$) within 0.7 Mpc of the cluster center.
 The fore/background counts are subtracted using five control fields. 
 Their results are more consistent with the steeper relation estimated in
 1995 by Rakos and Schombert than with the original one by Butcher and
 Oemler in 1984. The results are also consistent with ours. 
 Margoniner et al. (2001) extended their work to 295 Abell clusters
 and found $f_b$=(1.34$\pm$0.11)$\times$z-0.03 with a {\it r.m.s.}
 of 0.07, which is in agreement with our fitted function shown in Figure
 \ref{fig:bo}. 

 Ellingson et al. (2001) studied 15 CNOC1 clusters (Yee, Ellingson, \&
 Carlberg 1996) between z=0.18 and z=0.55. Since they used spectroscopically observed
 galaxies, they do not suffer from the fore/background
 correction (but see Diaferio et al. 2001). They used galaxies brighter than $M_{r}=-19.0$ within
 $r_{200}$ from the cluster center (with an average of 1.17$h^{-1}$
 Mpc). Their best fit shows $f_b\sim$0.15 at z=0.3. The scatter  in their Fig. 1 and our data are both substantial. 
 Thus, we can not
 conclude this is inconsistent with our results. 
 
 All of these  authors found considerable scatter
 in $f_b$ v.s. z plot as is seen in our Figure \ref{fig:bo}. 
 It is promising that our results are consistent with the previous
 work within the scatter, despite the differences in the radial coverage
 and magnitude ranges used. 

\subsection{Morphological Butcher-Oemler effect}

 In sections \ref{u-r}, \ref{exp}, and  \ref{Cin}, we found an
 increase in the fraction of late type galaxies selected by morphological parameters
 with increasing redshift --- as if the Butcher-Oemler effect is happening morphologically.
 Perhaps revealing this morphological Butcher-Oemler effect is the most striking result of this study.
 It suggests that the Butcher-Oemler blue galaxies
 change their morphology from late to early type at the same time that they
 change their color from blue to red. Although accurately quantifying the fraction of
 galaxies which experience the morphological Butcher-Oemler effect is
 difficult due to the considerable scatter in the data, our best-fit lines suggest 
 that $\sim$30\% of galaxies in clusters go through this transition
 between z=0.3 and z=0.02. 

  In previous work, Dressler et al. (1997)
 found a deficit of S0 galaxies in 10 intermediate  (z$\sim$0.5)
 clusters by classifying galaxy morphology in the HST image 
 by eye. They claimed that many S0s needed to be added to
 reach the fraction of S0s found in present clusters (Dressler 1980). 
 Couch et al. (1994, 1998) also found an indication of morphological
 transformation in the Butcher-Oemler  galaxies by studying three rich
 clusters at z=0.31.  Later, Fasano et al. (2000) showed that spiral
 galaxies are, in fact, turning into S0 galaxies 
 by observing nine clusters at intermediate redshifts and analyzing them
 together with higher redshift clusters in the literature. Their galaxy morphology was
 also based on eye classification. 
 Our SDSS data is taken using ground based telescopes with moderate
 seeing ($\sim$1.5''), and thus does not allow us to separate S0 galaxies
 from elliptical galaxies as the HST does.
 The advantage of our classification is its automated nature,
 which allows accurate reproducibility and quantification of
 systematic biases.
  In particular, it is easy to compute the completeness and contamination rate for
 the automated classification, based on simulations; for the present sample,
 the completeness and contamination rate of the parameters 
 are given in Shimasaku et al. (2001) and Strateva et
 al. (2001).  Furthermore,   an automated galaxy classification is easier to reproduce in
  future observational work and in detailed computer simulations.
 Although we can not distinguish S0s from ellipticals, the increase of blue
 fraction and increase of late type galaxies toward higher redshift is
 qualitatively consistent with the process of S0 production over the
 interval in cosmic time suggested by previous investigations.
 
 Various physical mechanisms could be the cause of the morphological
  and spectral Butcher-Oemler effects.
  Possible causes include ram pressure stripping of gas (Gunn \& Gott 1972, Farouki
 \& Shapiro 1980; Kent 1981, Abadi, Moore \& Bower 1999, Quilis, Moore \& Bower
 2000), galaxy infall (Bothun \& Dressler 1986, Abraham et al. 1996a, Ellingson et al. 2001), galaxy harassment via high speed impulsive encounters (Moore et al. 1996, 1999), cluster
 tidal forces (Byrd \& Valtonen 1990, Valluri 1993) which distort
 galaxies as they come close to the center, interaction/merging of
 galaxies (Icke 1985, Lavery \& Henry 1988, Bekki 1998), and removal \& consumption of the gas due to the cluster environment (Larson, Tinsley \& Caldwell 1980, Balogh et
 al. 2001, Bekki et al. 2002). Mamon (1992) and Makino \& Hut (1997) showed that
 interactions/mergers can occur in a rich cluster environment despite the
 high relative velocities. Shioya et al. (2002) showed that the
  truncation of star formation can  explain the decrease of S0 with
  increasing redshift. 
   It has been known that preheating of intergalactic medium
 can effect morphologies of galaxies by strangling the gas accretion (Mo
 \& Mao 2002; Oh \& Benson 2002). In fact, Finoguenov et al. (2003)
 found the filamentary gas in Coma cluster and predicted quiescent star
  formation in galaxy disks around the filament. 
  Although our results provide some important clues, pinpointing what
 processes are responsible in the morphological and spectral
 Butcher-Oemler effect is a more  difficult challenge. 

  Our results suggest that the cause will be one that affects both
 color and morphological appearance of galaxies at the same time. 
  Couch et al. (1998), Dressler et al. (1999) and Poggianti et al. (1999) found ``passive
 spirals'', which are galaxies with spiral morphology but without star
 formation.  They are probably belong to the same population as ``anemic
 spirals''  found by van den Bergh (1976). 
 The mechanism creating ``passive spirals'' or ``anemic
 spirals'', however, affects only the
 color of galaxies and, thus, 
 probably is not the main mechanism that accounts for the entire effect.
  The increase of morphologically late type galaxies toward higher
 redshifts at the same time as the 
 increase of blue galaxies is consistent with mechanisms which affect the gas supply
 (e.g. ram-pressure stripping, galaxy infall). However if the infalling
 rate of field galaxies (mostly blue/late type) is higher in the past,
 almost any of the mechanisms mentioned above can explain our
 observational results.
 Furthermore, although we
 discussed about cluster specific phenomena, it is also known that
 galaxies in the field region evolve as a function of redshift as well.
 (e.g. Hammmer et al. 1996; Lilly et al. 1996; Balogh et al. 1997,
 2002). The evolution of field galaxies needs to be compared
 with that of cluster galaxies further in detail.
 Therefore, it is still an open question what mechanism causes spectral
 and morphological evolution of cluster galaxies.


  The finding of a 30\% change of the fraction during the look back time
  of $\sim$3.5  Gyr could also give us an additional hint in finding an
  underlying   physical process.
  If the gas in spiral galaxies is removed very efficiently by some
 physical processes ($e.g.$ ram-pressure stripping) or consumed rapidly by
 star formation, the spiral arms will disappear after several disk
 rotation periods, $\sim$ 1 Gyr (Sellwood \& Carlberg 1984). 
 Interaction/merger processes are quicker than gas removal processes
  ($\sim$0.5 Gyr; Mihos 1995). 
 Moore et al.'s (1996) simulation showed that the galaxy harassment phase
 lasts for several Gyr. Kodama et al. (2001) used the phenomenological simulations to show that the timescale of the morphological transformation from spiral to S0 is 1$\sim$3 Gyr. 
 For spectral change,  Shioya et al. (2002) showed that a disk needs 2-3 Gyr after the removal  of gas (or truncation of star formation) to show a k spectrum. Poggianti
  et al. (1999) compared the spectral and morphological properties of
  cluster galaxies and suggested that the timescale of the
  morphological transition is longer than that of the spectral transition.
  This difference in timescale is interesting since if one process is
  significantly quicker than the other, we might be able to see the time
  difference in the decreases in the fraction of late type galaxies and
  blue galaxies, which will provide a strong constraint in the evolution history of the Butcher-Oemler galaxies.
 In Figure \ref{fig:bo}, we see a $\sim$30\%
 of change in both the photometric and morphological Butcher-Oemler effect
 between z=0.02 and z=0.30 ($\sim$3.5 Gyr). The
 scatter in our measurement, however, is considerable and our choices of 
 criteria between late and early type galaxies do not necessarily coincide
 with each other. It is thus not straightforward to convert the
  information to the time scale of the responsible physical process.
 In addition, to understand change in fraction of morphological and
  spectral late-type galaxies, the change in infalling rate of field
  galaxies needs to be understood as well.
 Since computer simulations have recently made dramatic progress,  
  in the near future it will become possible for
 state-of-the-art simulations to simulate both dynamical and
 spectral evolutions of cluster galaxies, plus infalling rate of field
  galaxies in order to compare the results with the
 observed trend. For example, such a simulation can be done by combining
  dynamical simulations (e.g. 
 Evrard 1991, Kauffmann et al. 1995, Bekki, Shioya \& Couch 2001,
  Vollmer et al. 2001, Bekki et al. 2002) with cluster
 phenomenological simulations (e.g. Abraham et al. 1996, Fujita
  1998,2001,  Balogh et al. 1999, Stevens, Acreman, \& Ponman 1999, 
 Balogh, Navarro, \& Morris 2000, Kodama \& Bower 2001).
 Figure \ref{fig:bo} in this work provides the interesting observational data
 to tackle with using such a simulation of cluster galaxy formation.

%
%
%
%
%
%
%

%
%

\subsection{Richness Dependence}

\begin{figure}[h]
\begin{center}
\includegraphics[scale=0.7]{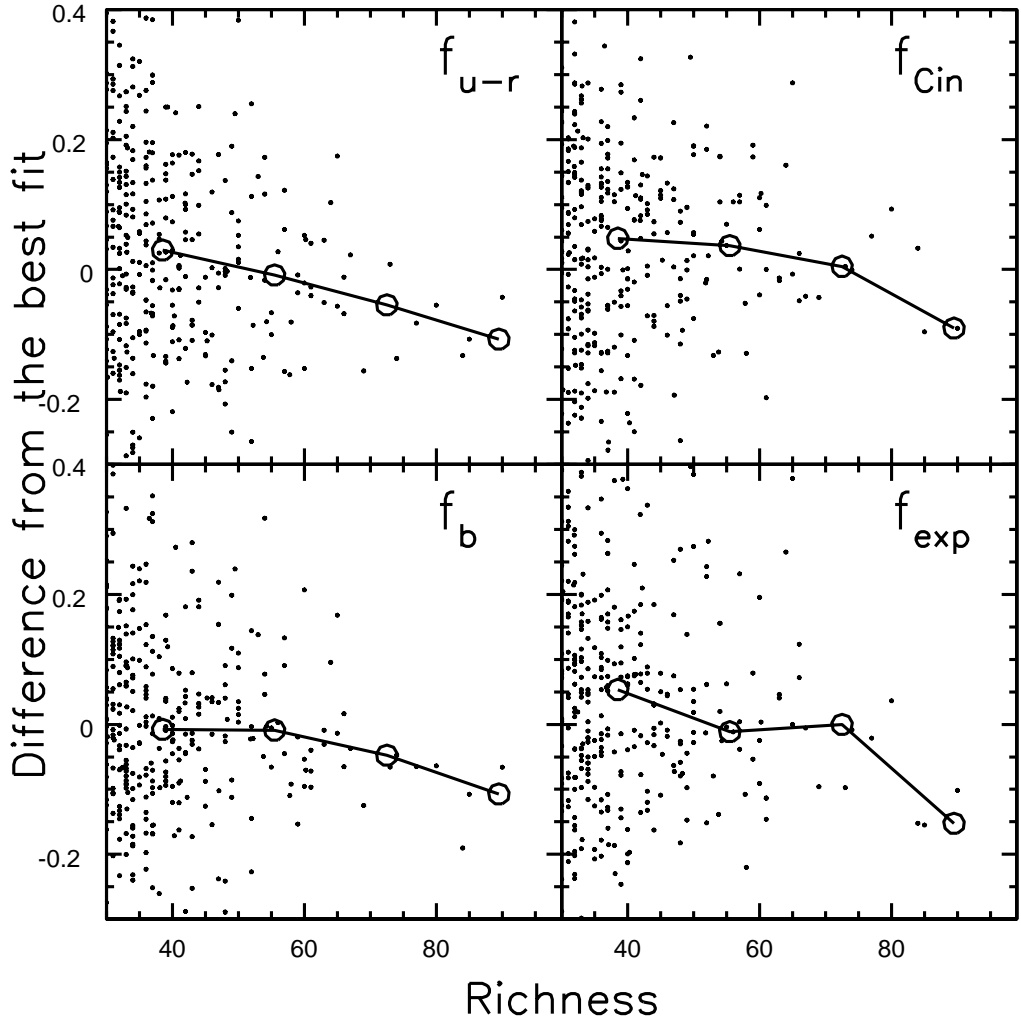}
\end{center}
\caption{
\label{fig:okamura_rich}
 The difference of the late type fractions from the best
 fit lines as a function of redshift are plotted against cluster richnesses. Solid lines and stars
 show the median values.
}
\end{figure}

\begin{figure}[h]
\begin{center}
\includegraphics[scale=0.7]{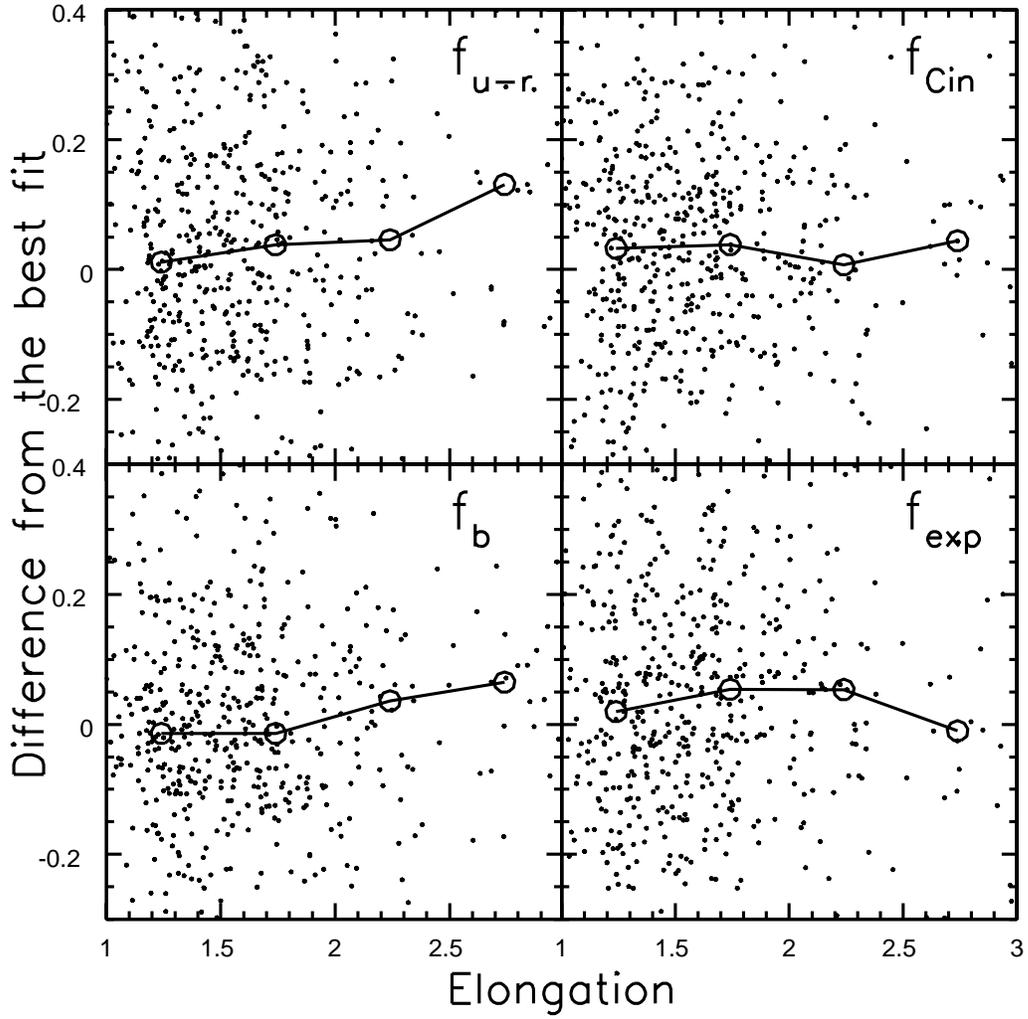}
\end{center}
\caption{
\label{fig:okamura_elong}
 The difference of late type fractions from the best
 fit lines as a function of redshift are plotted against cluster elongation, which was measured as
 a ratio of major axis to minor axis on a cluster detection map of Goto
 et al. (2002a).  Solid lines and stars show median values.
}
\end{figure}

\begin{figure}[h]
\begin{center}
\includegraphics[scale=0.7]{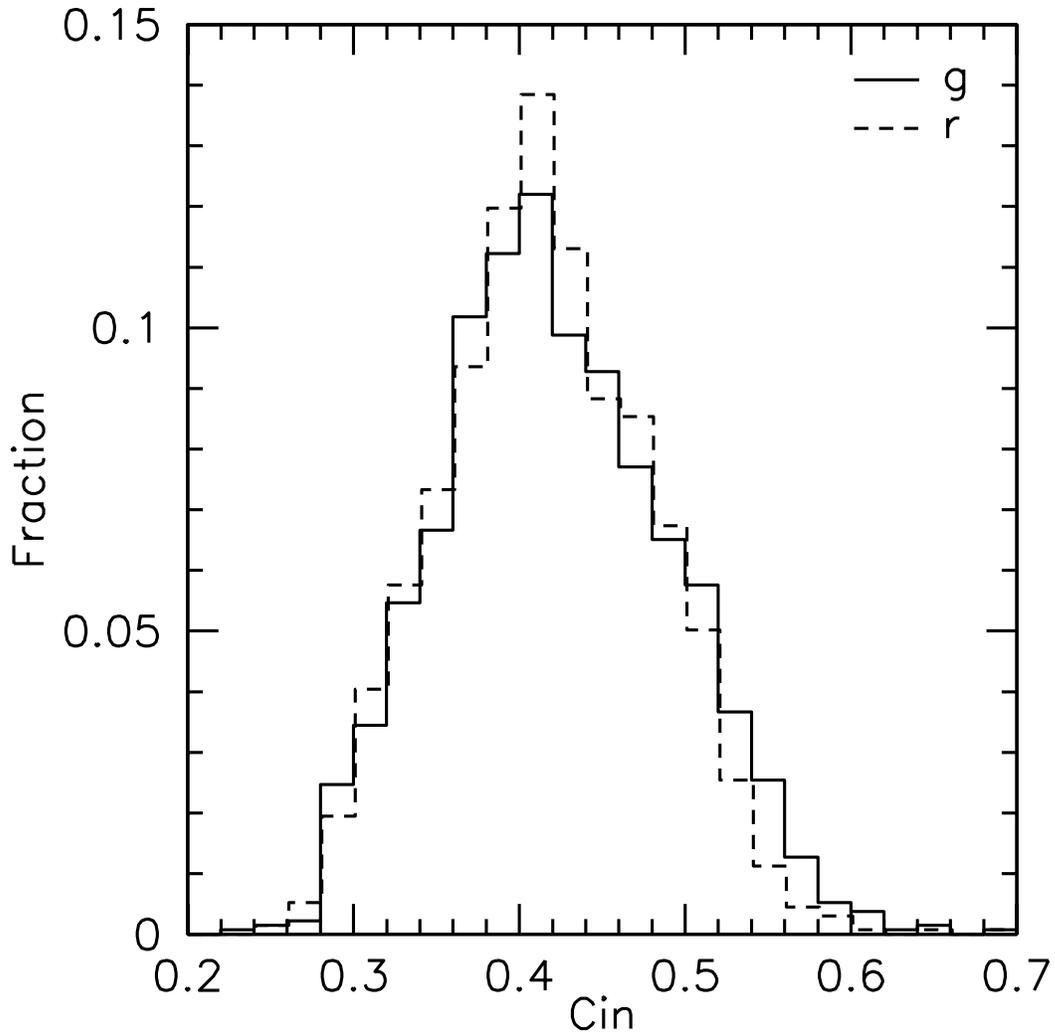}
\end{center}
\caption{
\label{fig:cin_g_r}
 The distribution of the inverse of the concentration index ($C_{in}$), defined
 as the ratio of Petrosian 50\% flux radius to Petrosian 90\% flux
 radius. The solid line shows the distribution of $C_{in}$ measured in
 the $g$ band image. The dashed line shows the distribution of $C_{in}$ measured in
 the $r$ band image. The difference between the $g$ band and $r$ band is
 marginal, assuring our usage of $r$ band $C_{in}$ in the right upper
 panel of Figure \ref{fig:bo} from $z$=0.02 to $z$=0.3. The
 statistics are summarized in Table \ref{tab:cin_g_r}.
}
\end{figure}

\begin{figure}[h]
\begin{center}
\includegraphics[scale=0.7]{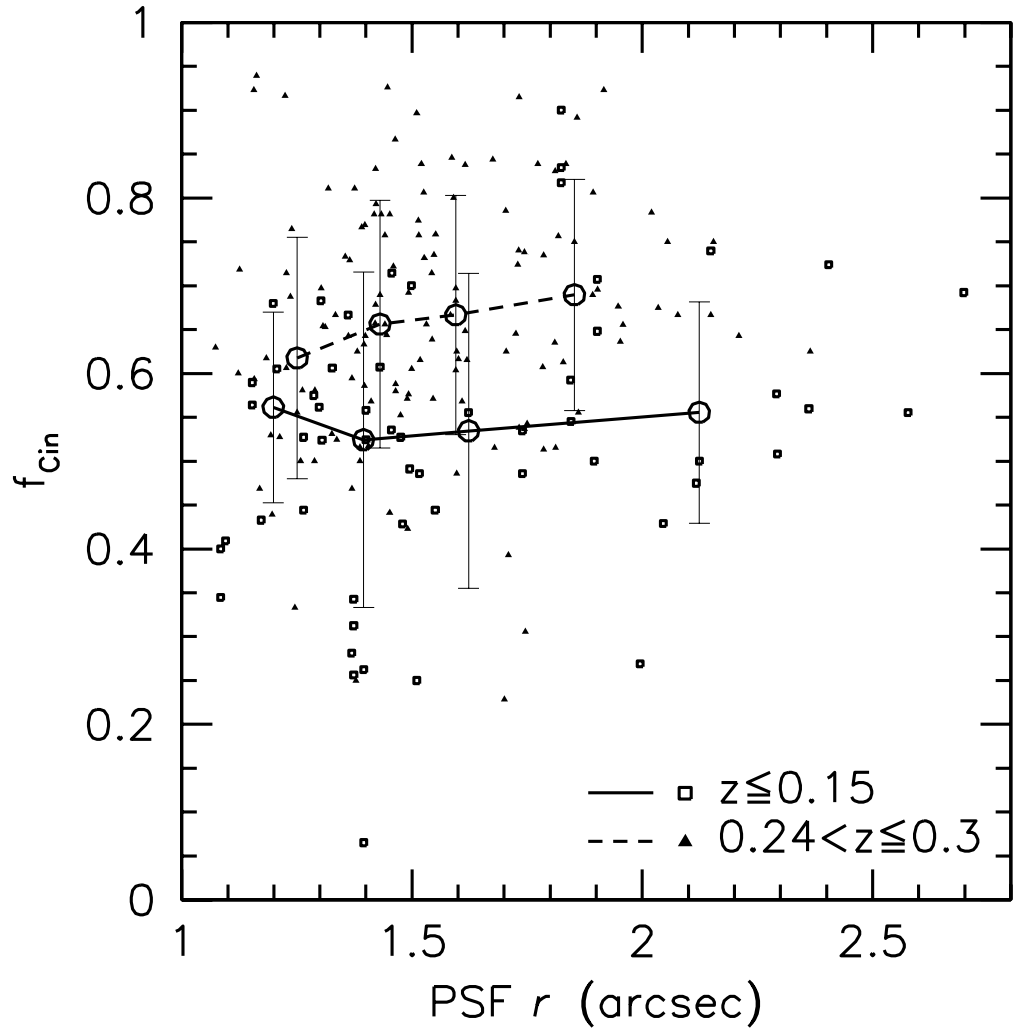}
\end{center}
\caption{
\label{fig:seeing_cin}
 The dependence of $f_{Cin}$ on seeing. Open squares and solid lines show
 the distribution and medians of low z clusters (z$\leq$0.15). Filled
 triangles and dashed lines show the distribution and medians of high z
 clusters (0.24$<$z$\leq$0.3). The median bins are chosen so that equal
 numbers of galaxies are included in each bin.
}
\end{figure}

\begin{figure}[h]
\begin{center}
\includegraphics[scale=0.7]{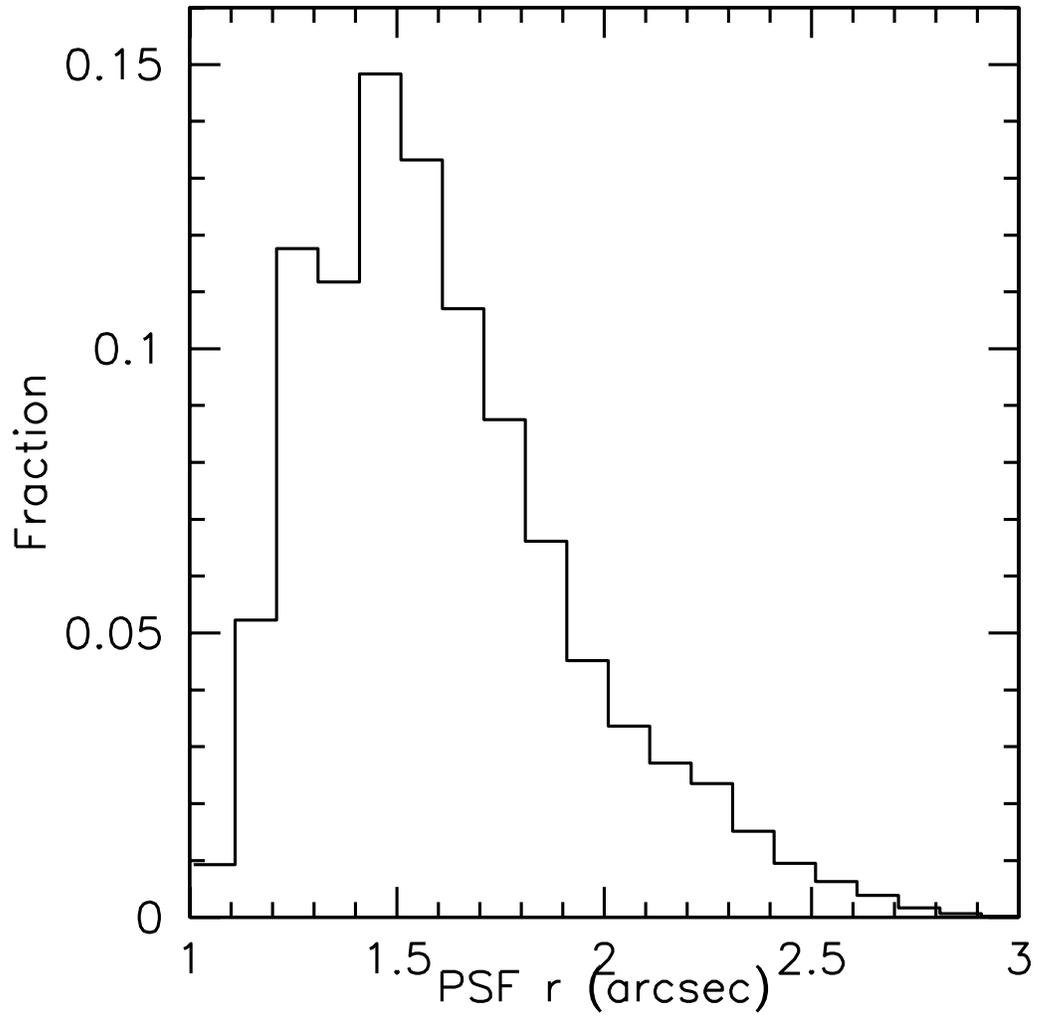}
\end{center}
\caption{
\label{fig:seeing_hist}
The seeing distribution of all galaxies brighter than $r$=21.5. 87\% of all
 galaxies have seeing better than 2.0 arcsec. 
}
\end{figure}

\begin{figure}[h]
\begin{center}
\includegraphics[scale=0.7]{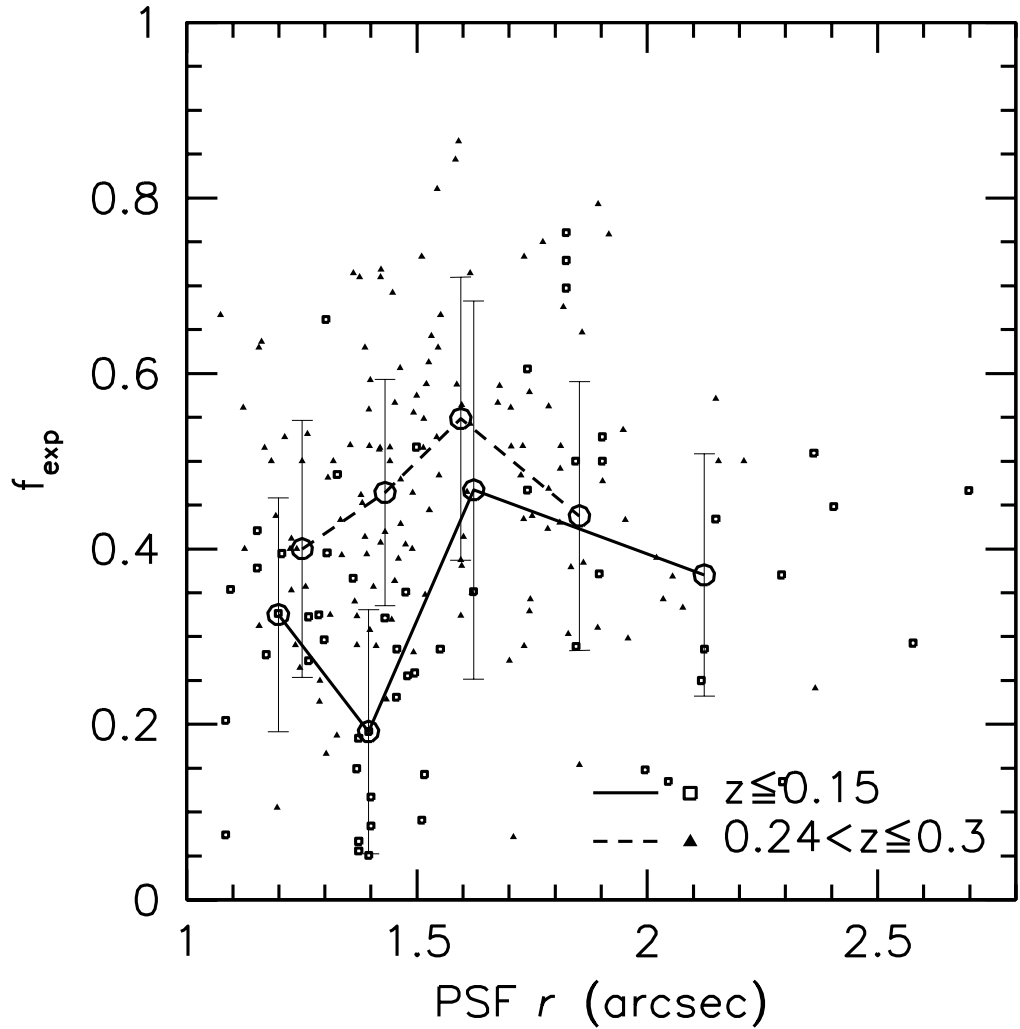}
\end{center}
\caption{
\label{fig:seeing_exp}
The dependence of $f_{exp}$ on seeing. Open squares and solid lines show
 the distribution and medians of low z clusters (z$\leq$0.15). Filled
 triangles and dashed lines show the distribution and medians of high z
 clusters (0.24$<$z$\leq$0.3). Median bins are chosen so that equal
 numbers of galaxies are included in each bin.
 1 $\sigma$ errors shown as vertical bars are more dominant.
 There is no significant trend with seeing.
}
\end{figure}

\begin{figure}[h]
\begin{center}
\includegraphics[scale=0.7]{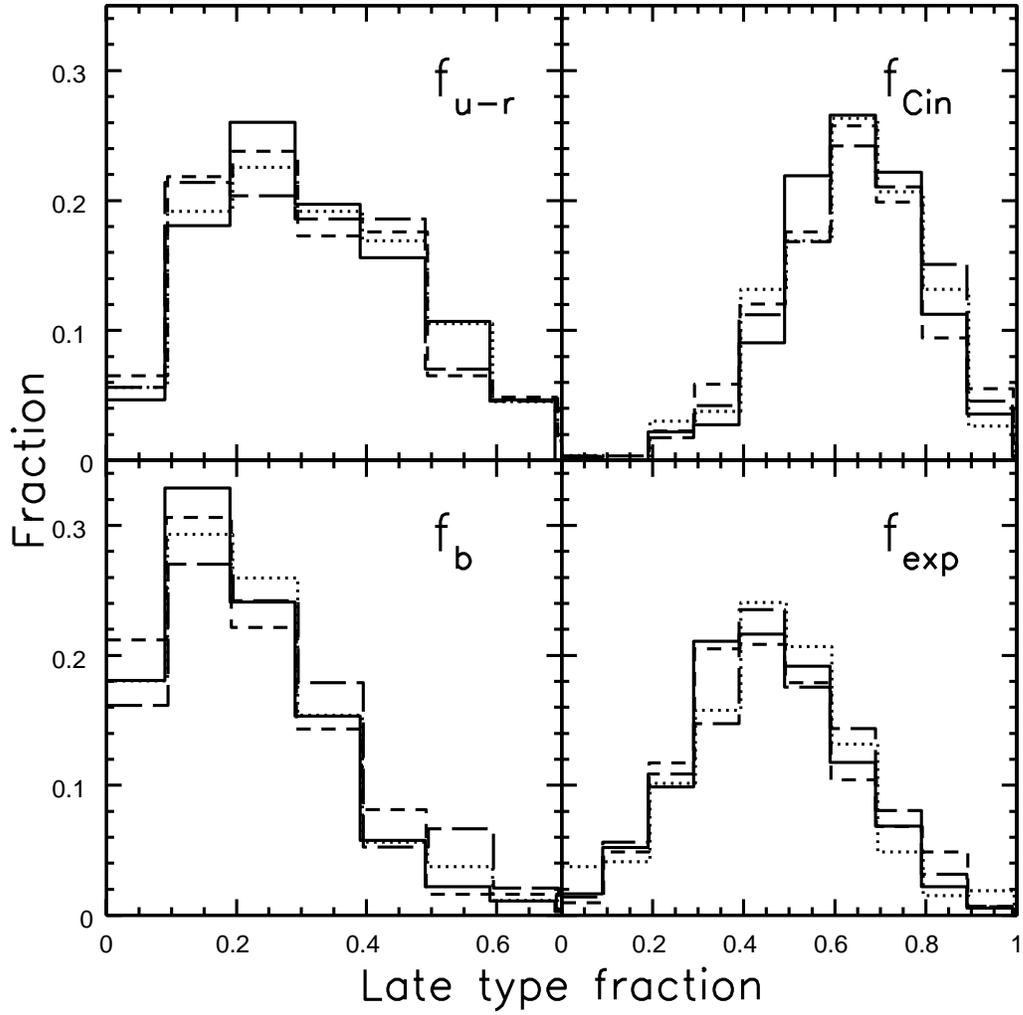}
\end{center}
\caption{
\label{fig:bg_test}
 Various systematic tests. Solid lines show distributions for
 2.1-2.21 Mpc annular fore/background subtraction. Dashed lines show
 distributions for a global background subtraction. Dotted lines show
 distributions using 0.7/(1+z) Mpc radius assuming a standard cold dark matter cosmology. 
 Long dashed lines show distributions using the brightest galaxy position as a cluster center. 
  In none of the cases does a Kolomogorov-Smirnov test show
 significant difference between the distributions (significance to be different is less than 26\% in all cases).
}
\end{figure}

\begin{figure}[h]
\begin{center}
\includegraphics[scale=0.7]{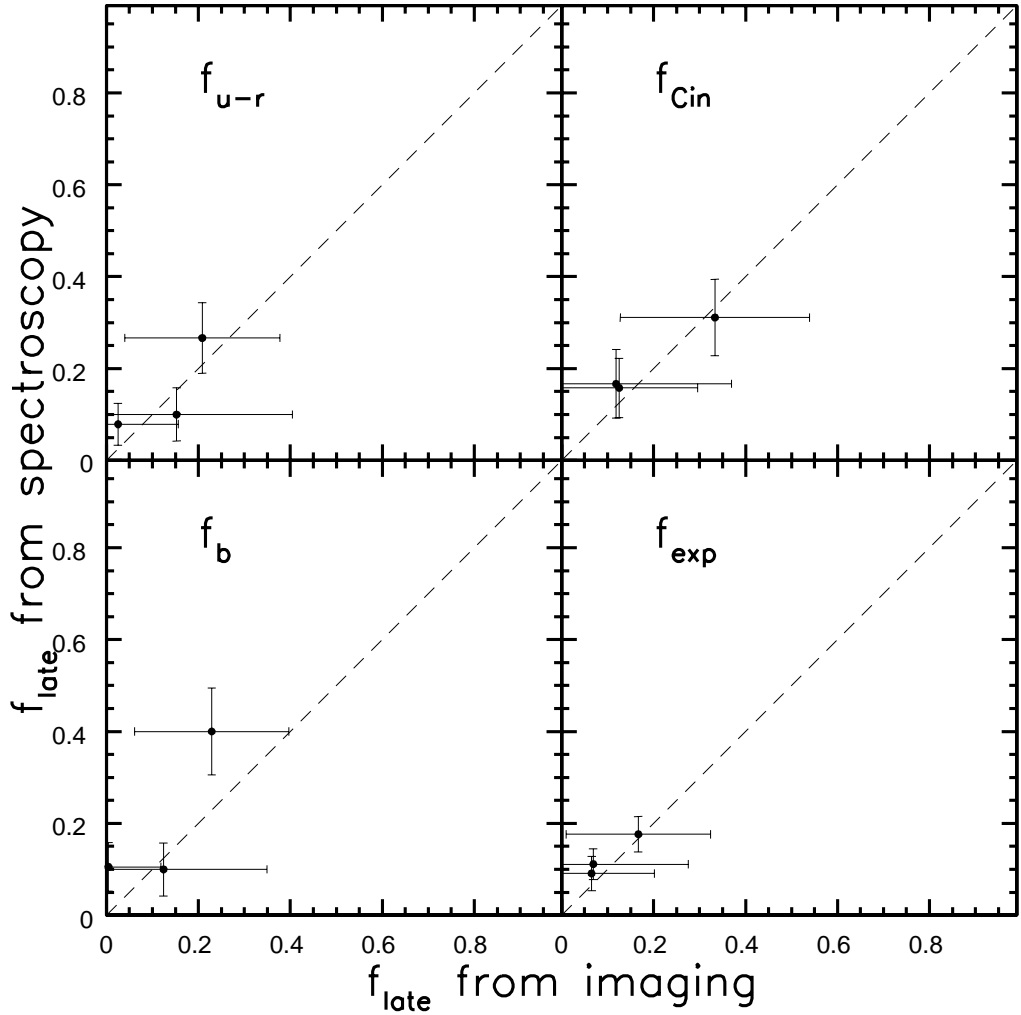}
\end{center}
\caption{
\label{fig:comparison_with_spectroscopy}
Comparison with late-type fraction from spectroscopy. 
 Late-type fractions measured using spectroscopic data are plotted
 against that from imaging data for three clusters with z$<$0.06 (ABELL 295, RXC
 J0114.9+0024, and  ABELL 957). Dashed lines are drawn to guide
 eyes. All points agree with each other within the error. 
}
\end{figure}

  In Section \ref{scatter}, we observe the tendency of richer clusters
 to have smaller fractions of late type galaxies, by measuring the
 residuals from the best fit relations as a function of cluster
 richness. Our result is consistent with Margoniner et al. (2001), who
 used a similar optical richness to find that poorer clusters tend to
 have larger blue fractions than 
 richer clusters at the same redshift. Figure \ref{fig:okamura_rich},
 however, still shows a significant amount of scatter, which might be
 suggesting the existence of another factor in determining the blue
 fraction in addition to redshift and richness.   
 The dependence of the late type fraction on cluster richness, however, provides
 another hint on the underlying physical processes. Since ram pressure
 is stronger in clusters with higher temperature at the same gas density, Fujita \&
 Nagashima(1999) theoretically predicted that if ram pressure is the only
 mechanism responsible for the evolution of galaxies in clusters, the 
 fraction of blue galaxies will always be higher in lower X-ray luminosity
 clusters, which usually have low temperatures. Our data shown in Figure
 \ref{fig:okamura_rich} are consistent with the prediction from their ram
 pressure stripping model. Although
 our richness is from numbers of galaxies in optical imaging data, it is reasonable to assume it correlates well with X-ray luminosity (Bahcall 1977;
 Bower et al. 1994). Then, the optical
 richness can be related to the gas temperature through the well known $L_{X}-T$
 relation (Mitchell, Ives, and Culhane 1977; Henry \& Arnaud 1991; Edge
 \& Stewart 1991; David et al. 1993; White, Jones, and Forman 1997; Allen
 \& Fabian 1998; Markevitch 1998; Arnaud \& Evrard 1999; Jones \& Forman
 1999; Reichart, Castander, \& Nichol 1999; Wu, Xue, and Fang 1999; Xue
 \& Wu 2000; and see the references therein).
 In a simple estimation, ram pressure is proportional to $\rho
 v^2$. $L_X$ is proportional to $\rho^2$. From the virial theorem, $v^2\propto T$.
 The $L_{X}-T$ relation studied by Xue \& Wu (2000) is $L_X\propto T^{2.8}$.
 Therefore, ram pressure is proportional to $\sim L_X^{0.86}$. 
 Combined with an assumption that optical richness scales with X-ray
 luminosity (see, e.g. Bahcall 1974, Jones \& Forman 1978, Bower et
 al. 1994, and Miller et al. in preparation),  
 Figure \ref{fig:okamura_rich} provides another hint that ram
 pressure stripping induces the evolution of cluster galaxies.

 In the literature, however, the dependence of blue fractions on cluster richness has been controversial. Bahcall (1977) studied 14 X-ray clusters and found that
 the fraction of spiral galaxies decreases with increasing X-ray
 luminosity.  
 Lea \& Henry(1988) observed 14 clusters in X-ray and found that 
 the percentage of blue objects in the clusters seems to increase with the X-ray luminosity.
 On the other hand, Fairley et al. (2002) studied eight X-ray selected
 clusters and did not find any dependence of blue fractions on X-ray
 luminosities. Balogh et al. (2002) studied 10 clusters at z=0.25 with low X-ray
 luminosity and found similar morphological and spectral properties of
 galaxies compared with clusters with high X-ray luminosity (Balogh et
 al. 1997). 
 In all cases, the results were based on a small sample
 of clusters.  We also point out that although our results are consistent with
 a ram-pressure stripping model, there is a possibility that the other
 mechanisms could explain the phenomena. For example, richer clusters
 might have higher rate of merger/interaction due to their higher galaxy
 density. The same argument holds true for galaxy harassment. Thus, more
 study is needed to conclude about the physical mechanism causing the phenomena.
 In the near future, confirming the richness
 dependence using X-ray luminosities or velocity dispersions with a larger
 sample of clusters would offer us more insight on the subject.

\section{Conclusions}\label{conclusion}

  In this paper, we have investigated the fraction of late type galaxies in
  four different ways using one of the largest, most uniform samples of
  514 clusters between 0.02$\leq$z$\leq$0.3 from the SDSS Cut \& Enhance galaxy cluster catalog.
  All the clusters selected here have more than 25 member galaxies within
  0.7 Mpc and between $M_r=-24.0$ and $M_r=-19.44$ after statistical
  local fore/background 
  subtraction. The following four different ways to estimate the fractions of late
  type galaxies are adopted: restframe $g-r$ color (a classical Butcher-Oemler estimator), $u-r$ color, concentration  index and de Vaucouleur/exponential profile fit. 
 The last three parameters are known to correlate well with galaxy
 morphologies (Shimasaku et al. 2001, Strateva et al. 2001).
 In all four cases, we observe an
  increase of the fraction of late type galaxies toward higher redshift
  with a significance of more than 99.99\% (Table \ref{tab:correlation}).   
 We draw the following conclusion from this work.

 1. We confirm the presence of the Butcher-Oemler effect using $g-r$ color. The Butcher-Oemler effect is real and exists in our sample clusters as seen in the lower left panel of Figure \ref{fig:bo}. The slope of the increase is consistent with previous work although the scatter in the blue fraction is considerable. 
 Previous work also noted a large scatter in the fraction of blue galaxies. 
 The fraction of late type galaxies also shows a similar increase when
 we use a $u-r$ color cut. 


 2. We observe a morphological Butcher-Oemler effect as an
 increase of late type galaxies toward higher redshift, using pure morphological
 parameters such as a concentration parameter and de
 Vaucouleur/exponential profile fit. The rates of increase
 are consistent with previous work on the spiral to S0
 transition, albeit with  considerable scatter (Figure \ref{fig:bo}).
 The increase is also in agreement with the original Butcher-Oemler effect
 from $g-r$ color. Our results are consistent with the evolutionary
 scenario proposed by Dressler et al. (1997), Smail et al. (1997), Couch
 et al. (1998), and Kodama \& Smail (2002), 
 in which there is a progressive morphological conversion in
 clusters from spirals into E/S0's.  

 3. We find a slight tendency for richer clusters to have lower values of
    the late type fraction (Figure \ref{fig:okamura_rich}).  This trend
    agrees with the ram pressure stripping model proposed by Bahcall (1977)
    and Fujita et al. (1999), in
    which galaxies in richer clusters are more affected by ram pressure due to
    their high temperature. 
%


 Although our results 1,2, and 3 are all consistent with a ram-pressure
 stripping model, there still remains a possibility that other physical
 mechanisms are  responsible for the evolution of cluster galaxies.
 Thus, further study is needed both theoretically and observationally to
 reveal the underlying physical mechanism responsible for the evolution
 of cluster galaxies.
 Since this work is based on only 5\% of the whole SDSS data, an
 increase in the data will improve the statistical accuracy as the SDSS
 proceeds. Extending the work to higher redshifts using 4-8 m class
 telescopes will offer more insight on the origin and evolution of
 cluster galaxies.

%
%

\bigskip


 We are grateful to Nell Hana Hoffman, Ricardo Colon, Michael L. Balogh,
 A. Kathy Romer and
 Robert C. Nichol for valuable comments, which contributed to improve
 the paper.
 We thank anonymous referee for valuable comments, which improved the
 paper significantly.   
 T.G. acknowledges financial support from the Japan Society for the
 Promotion of Science (JSPS) through JSPS Research Fellowships for Young
 Scientists. 

 Funding for the creation and distribution of the SDSS Archive has been provided by the Alfred P. Sloan Foundation, the Participating Institutions, the National Aeronautics and Space Administration, the National Science Foundation, the U.S. Department of Energy, the Japanese Monbukagakusho, and the Max Planck Society. The SDSS Web site is http://www.sdss.org/.

 The SDSS is managed by the Astrophysical Research Consortium (ARC) for the Participating Institutions. The Participating Institutions are The University of Chicago, Fermilab, the Institute for Advanced Study, the Japan Participation Group, The Johns Hopkins University, Los Alamos National Laboratory, the Max-Planck-Institute for Astronomy (MPIA), the Max-Planck-Institute for Astrophysics (MPA), New Mexico State University, Princeton University, the United States Naval Observatory, and the University of Washington.

\appendix

\section{Varying Radius}\label{varying}

\begin{figure}[h]
\begin{center}
\includegraphics[scale=0.7]{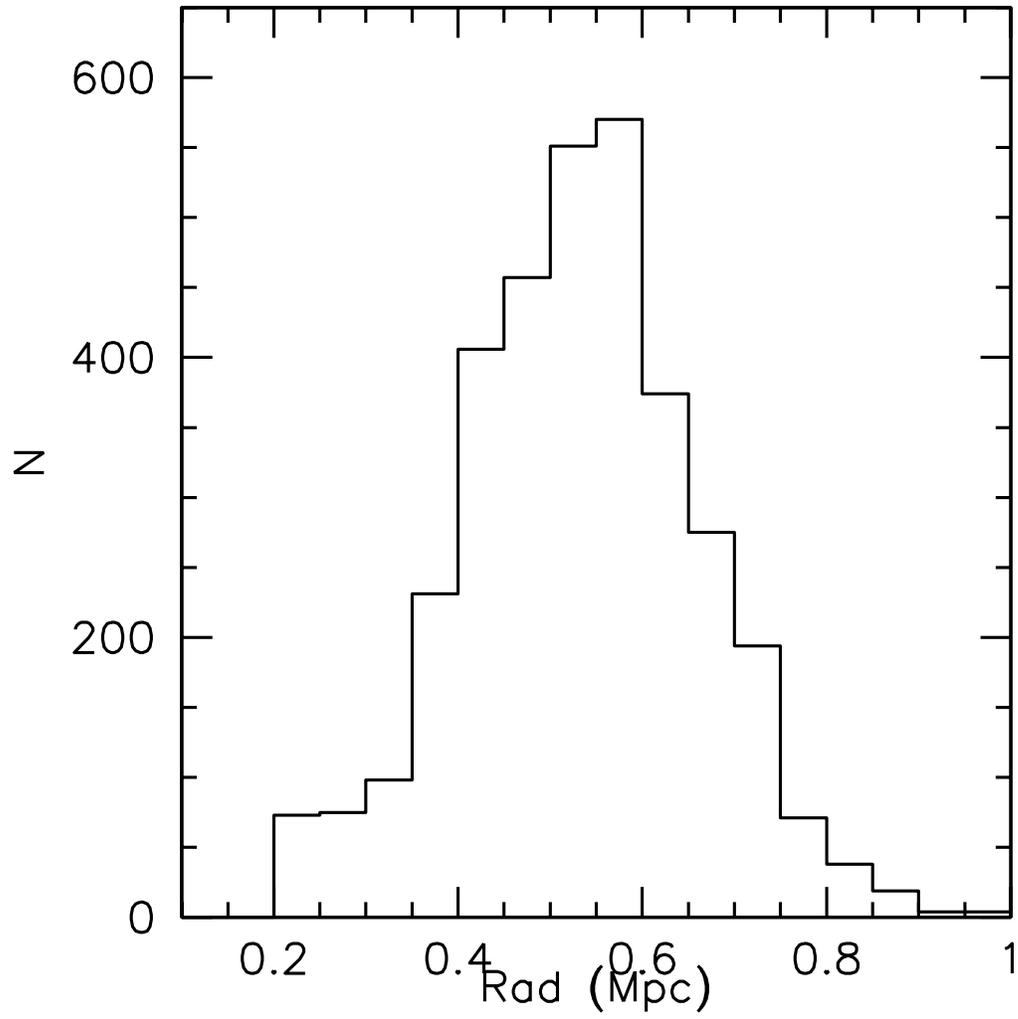}
\end{center}
\caption{
 Distribution of varying radius to measure blue/spiral fractions. It has
 a peak at 0.7 Mpc.
}\label{fig:new_rad}
\end{figure}

\begin{figure}[h]
\begin{center}
\includegraphics[scale=0.7]{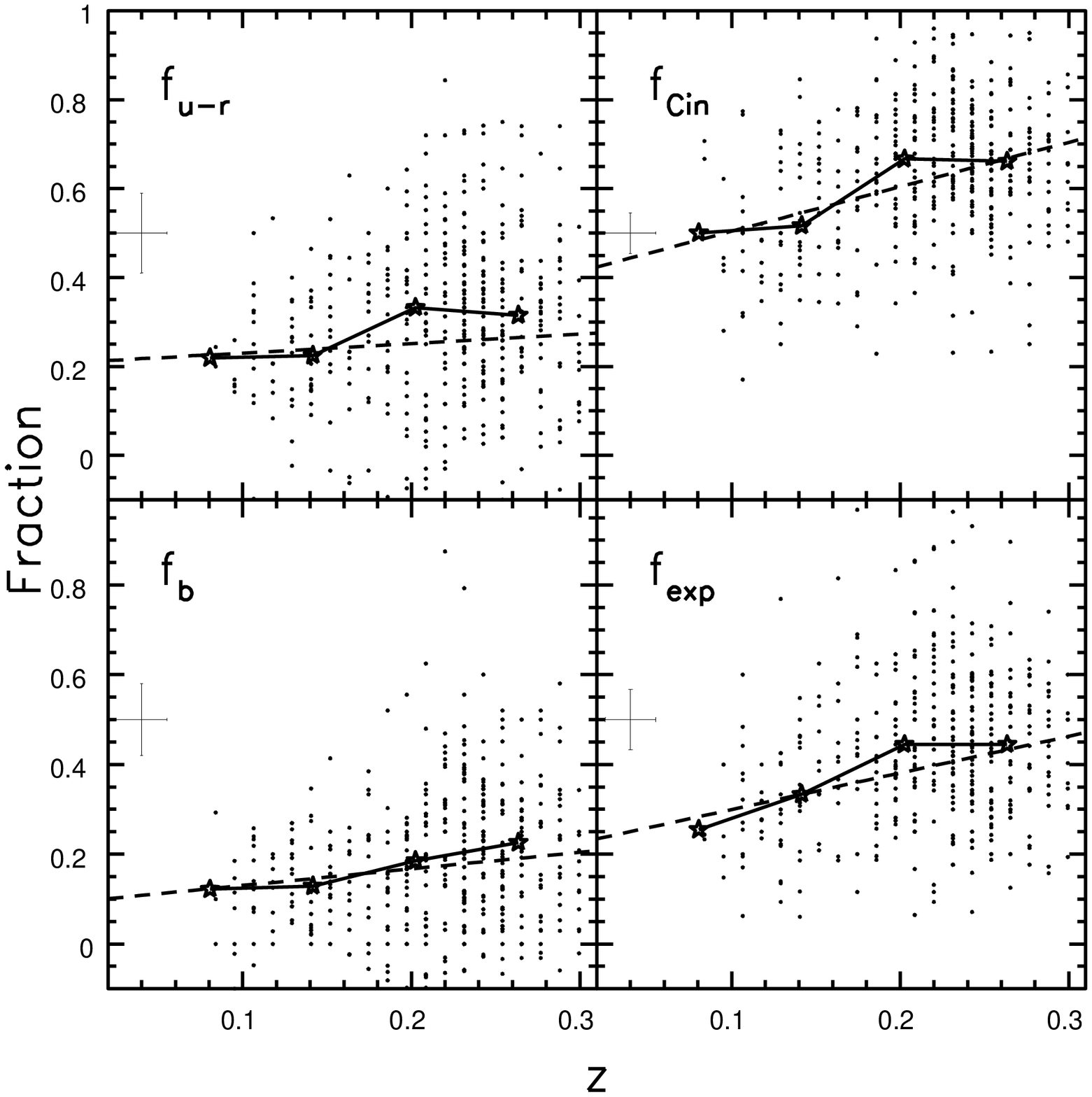}
\end{center}
\caption{
 The same as figure \ref{fig:bo}, but measured with varying radius.
}\label{fig:bo_new}
\end{figure}

\begin{figure}[h]
\begin{center}
\includegraphics[scale=0.7]{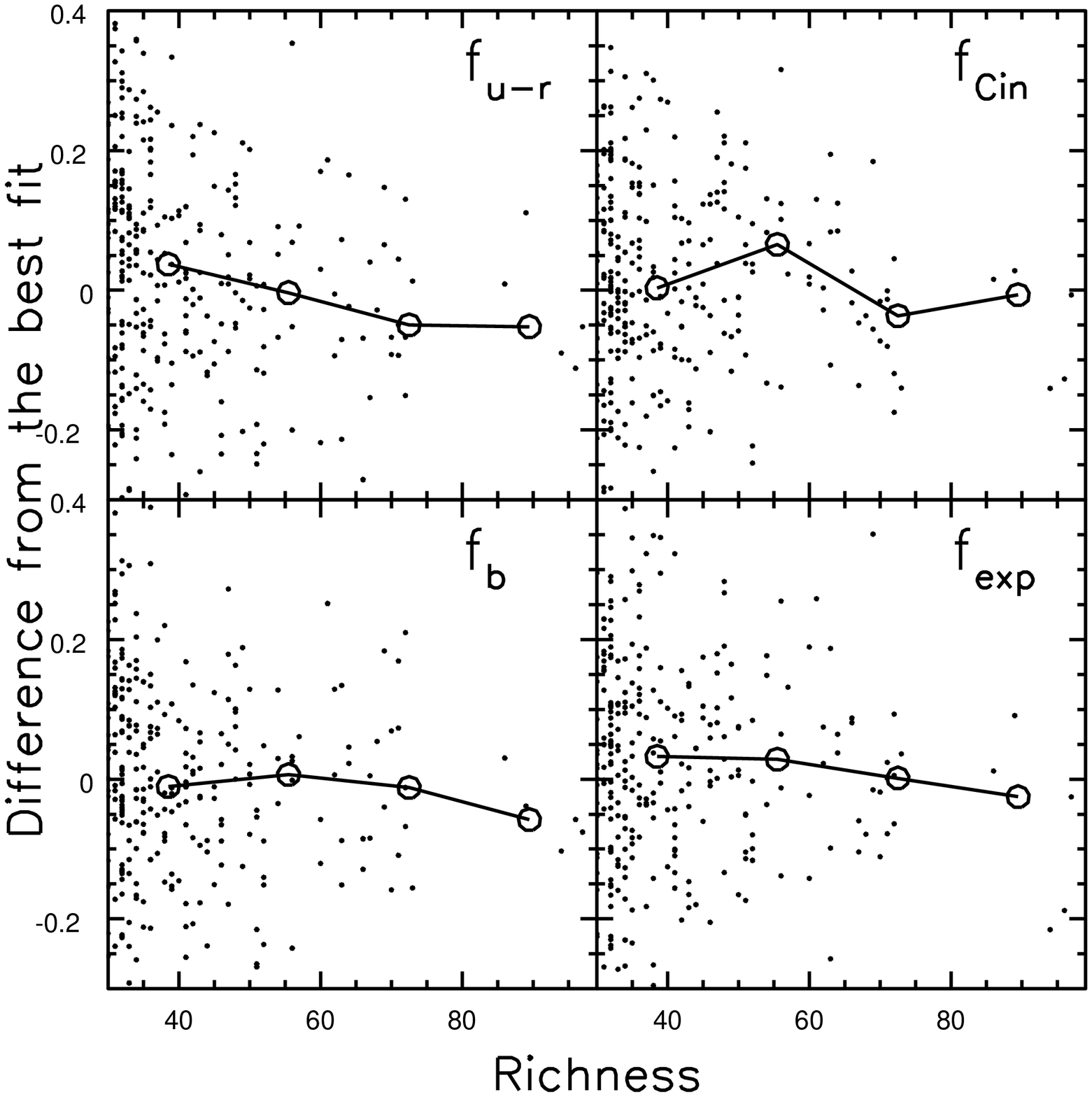}
\end{center}
\caption{
 The same as figure  \ref{fig:okamura_rich}, but measured with varying radius.
}\label{fig:okamura_rich_new}
\end{figure}

\begin{figure}[h]
\begin{center}
\includegraphics[scale=0.7]{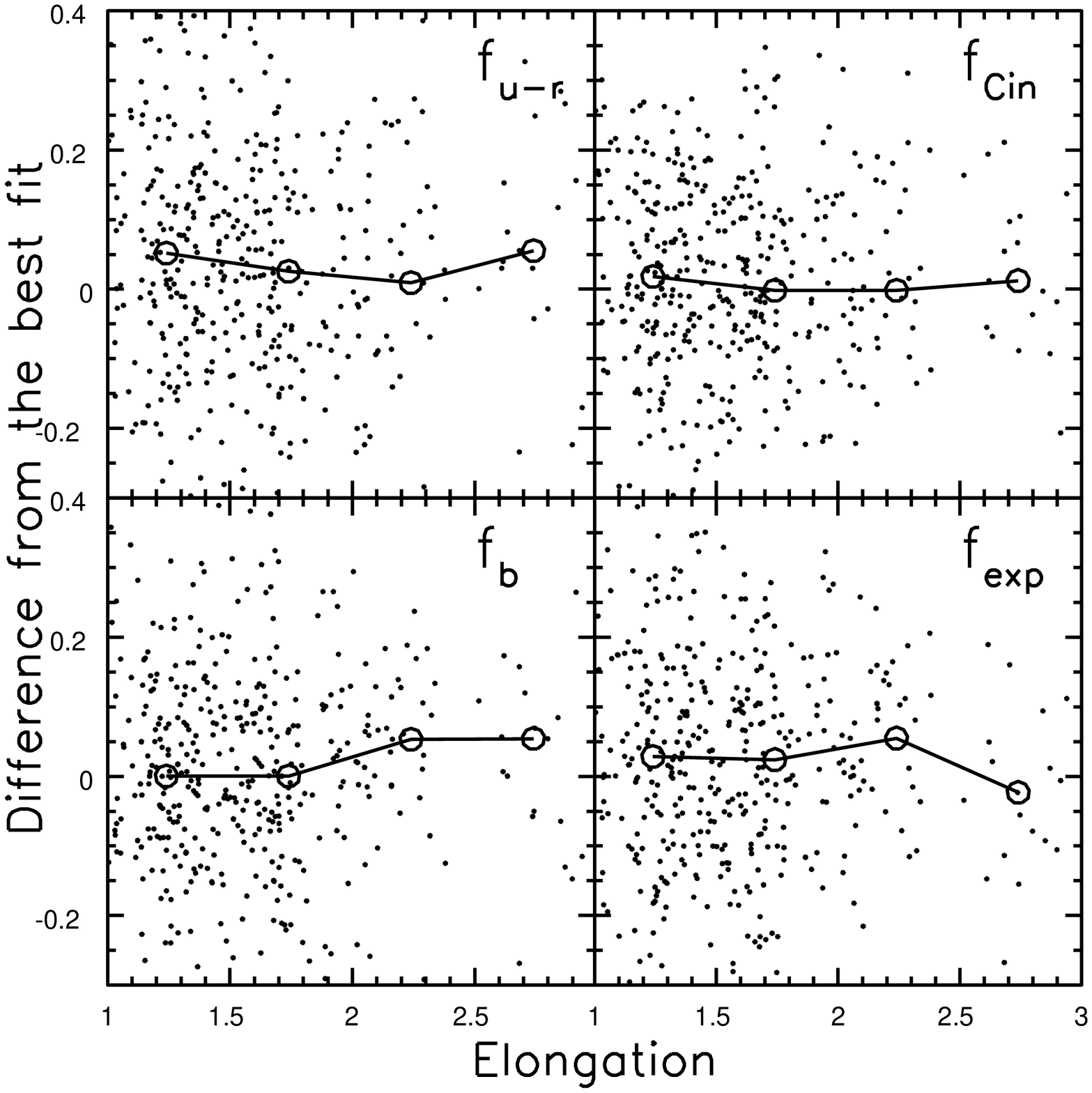}
\end{center}
\caption{
  The same as figure \ref{fig:okamura_elong}, but measured with varying radius.
}\label{fig:okamura_elong_new}
\end{figure}

 In Section \ref{method}, we used a fixed 0.7 Mpc radius to measure
 blue/spiral fractions among cluster galaxies since it was difficult to
 measure virial radius for relatively poor clusters in our sample from the
 SDSS imaging data. 
 However it is known that
 virial radius changes according to cluster richness; i.e. richer
 clusters have larger virial radius than poorer clusters. Therefore
 using a fixed radius could bring some bias associated with cluster
 richness. In this section we try to rectify this problem using cluster
 richness to calculate virial radius under a simple assumption. 
 We assume that our cluster richness (number of galaxies between $M_{r^*}=-24.0$ and  $M_r=-19.44$
 within 0.7 Mpc 
 after fore/background subtraction) is proportional to volume of a cluster, and
 therefore proportional to $radius^3$. Since richness is a relatively
 easy parameter to measure from the imaging data, we use the following
 equation to calculate radius for each cluster.
 
  \begin{equation} 
    radius = 0.7\times(Richness/32)^{1/3}
 \end{equation}

 where median richness of our sample cluster is 32. The coefficient of
 the equation is adjusted so that median clusters in our sample have
 radius of 0.7 Mpc, which corresponds to the mean radius used in Butcher et
 al. (1978, 1984) and Margoniner et al. (2000, 2001). The distribution of
 radius calculated in this way is presented in figure \ref{fig:new_rad}. As expected
 it has a peak at 0.7 Mpc. 
 Using this varying radius, we re-calculated all figures in Section
 \ref{method}. Sample clusters are still required to have more than 25
 galaxies after fore/background subtraction within the new
 radius. Therefore the number of sample clusters are somewhat reduced to
 413 clusters. 
 Results are presented in figures
 \ref{fig:bo_new}-\ref{fig:okamura_elong_new}. 
 Reassuringly, all 
 figures have the same trend as presented in Section
 \ref{method}. Therefore the discussion in Section \ref{discussion}
 still holds.  Although it is ideal
 to use virial radius to measure blue/spiral fractions of clusters, we
 regard that our analysis using fixed 0.7 Mpc radius is not hampered to
 the extend where our main conclusions change.

 \section{Errors on Blue/Late Type Fractions}\label{error}

 In this section we summarize how we derived eq. (\ref{frac_err}) to
 calculate errors on blue/late type fractions. 
  As a starting point, we assume the following.

 \begin{itemize}
  \item Number of galaxies in a cluster region ($N_{c+f}^{all}$) follows
	Poisson statistics.
  \item Number of galaxies in a certain area of field region ($N_{f}^{all}$) follows
	Poisson statistics. 
  \item $N_{c+f}^{*}$ and $N_{f}^{*}$ are independent of each other.
  \item Number of blue/late type galaxies in a cluster region ($N_{c+f}^{late}$) is
	strongly correlated with number of all galaxies in that region
	($N_{c+f}^{all}$).
  \item  Number of blue/late type galaxies in a certain field region ($N_{f}^{late}$) is
	strongly correlated with number of all galaxies in that region
	($N_{f}^{all}$).
 \end{itemize}

 And we clarify the definition of our notation.
 In this appendix, $\delta A $ means a deviation of a sampled value
 from an expectation value, $E(A)$. \\
 $\delta A$=$A-E(A)$, where A is each data value.\\
 $E(A)$ and $\delta A$ satisfy the following relations.\\
 $E(\delta A)$=0, $E(\delta A)^2$=$\sigma^2$,
 and if A and B are independent, $E(\delta A \delta B)$=0.
 Note that the equation \ref{frac_err} is not a deviation of
 a single sample but the expectation value estimated from the sample,
 and should be written as $E(\delta f_{late}^2)$ if we write rigidly.

 Under these assumptions, the error of late type fraction, 
 $\delta f_{late}$, become
 \begin{equation} \label{xy}
 \delta f_{late}^2 =   (X/Y)^2 \times ((\delta X^2/X^2) + (\delta Y^2/Y^2) -2(\delta X \delta Y/XY)),
 \end{equation}
 where $X$=$N_{c+f}^{late}-N_{f}^{late}$,
 $Y$=$N_{c+f}^{all}-N_{f}^{all}$, and $f_{late}$ = $X/Y$. 

Since $N_{c+f}^{all}$ and $N_{f}^{all}$ follow Poisson statistics,
         and they are independent. 
         \begin{equation}
	       E(\delta Y^2) = N_{c+f}^{all}+N_{f}^{all}\end{equation} 
         Similarly, when  $N_{c+f}^{all}$ follows Poisson statistics,
         $N_{c+f}^{late}$ also follows  Poisson statistics
         since  $N_{c+f}^{late} \sim N_{c+f}^{all}\times f_{late}$.
	 Therefore,
 	 \begin{equation}
	 E(\delta X^2) = N_{c+f}^{late}+N_{f}^{late}.\end{equation}
 
           Deriving the cross term at the end of the equation is not
             so straightforward. 
	     The cross term is expanded as 
\begin{equation}
               \delta X \delta Y 
=  \delta(N_{c+f}^{late}-N_{f}^{late})\delta(N_{c+f}^{all}-N_{f}^{all})\end{equation}\begin{equation}
	     =(\delta(N_{c+f}^{late})-\delta(N_{f}^{late}))(\delta(N_{c+f}^{all})-\delta(N_{f}^{all}))\\\end{equation}\begin{equation}
	     =\delta(N_{c+f}^{late})\delta(N_{c+f}^{all})
	     -\delta(N_{f}^{late})\delta(N_{c+f}^{all})
	     -\delta(N_{c+f}^{late})\delta(N_{f}^{all})
	     +\delta(N_{f}^{late})\delta(N_{f}^{all})\end{equation}

	     Since we assume $N_{f}^{late}$ and $N_{c+f}^{all}$,
	     $N_{c+f}^{all}$ and $N_{f}^{late}$ are both independent,
	 \begin{equation}
	 E(-\delta(N_{f}^{late})\delta(N_{c+f}^{all}))=0 \end{equation}

	     and
\begin{equation}
E(-\delta(N_{c+f}^{late})\delta(N_{f}^{all}))=0. \end{equation}

	     Therefore,  
\begin{equation} 	      
  E(\delta X \delta Y) =
  E(\delta(N_{c+f}^{late})\delta(N_{c+f}^{all})+\delta(N_{f}^{late})\delta(N_{f}^{all}))\\
\end{equation}

  Since we assume that $N_{c+f}^{late}$ and $N_{c+f}^{all}$, or
  $N_{f}^{late}$ and $N_{f}^{all}$ strongly correlate,
	 we can approximate that\\
\begin{equation}
 	  E(\delta(N_{c+f}^{late})\delta(N_{c+f}^{all}))=
	  \sigma(N_{c+f}^{late})\sigma(N_{c+f}^{all}) =
\sqrt{N_{c+f}^{late}}\sqrt{N_{c+f}^{all}} \end{equation}

	  and,\\
\begin{equation}\label{field_approximation}
 	  	  E(\delta(N_{f}^{late})\delta(N_{f}^{all}))=
	  \sigma(N_{f}^{late})\sigma(N_{f}^{all}) =
\sqrt{N_{f}^{late}}\sqrt{N_{f}^{all}}\\ \end{equation}

	 Therefore we obtain,
	 \begin{equation}\label{deltaxy}
 	   E(\delta X \delta Y) =  \sqrt{N_{c+f}^{late}}\sqrt{N_{c+f}^{all}} +\sqrt{N_{f}^{late}}\sqrt{N_{f}^{all}}
	  \end{equation}

            By substituting eq. (\ref{deltaxy}) for $\delta X \delta Y$
            in eq. (\ref{xy}) , we derive
            eq. (\ref{frac_err}).

	However this is not the only way to estimate the error.  Actually, the
  correlation between  $N_{f}^{late}$ and $N_{f}^{all}$ is not so
  obvious since late type fraction in the field and that in the cluster
  region might be different. Although we regard the difference is so small
  that we can assume the eq.(\ref{field_approximation}),
  if we assume that $N_{f}^{late}$, $N_{f}^{early} (= N_{f}^{all} - N_{f}^{late})$,
  $N_{c+f}^{late}$, and $N_{c+f}^{early} (= N_{c+f}^{all} - N_{c+f}^{late})$ 
  are independent, we derive,
\begin{equation}
 	  E(\delta(N_{c+f}^{late})\delta(N_{c+f}^{all}))=
          E(\delta(N_{c+f}^{late})(\delta(N_{c+f}^{late})+\delta(N_{c+f}^{early})))=
	  \sigma(N_{c+f}^{late})^2 = N_{c+f}^{late} \end{equation}
	  and,\\
\begin{equation}
 	  	  \delta(N_{f}^{late})\delta(N_{f}^{all})=
	  \sigma(N_{f}^{late})^2 = N_{f}^{late}. \end{equation}
Then, the expectation value of $\delta X \delta Y$  becomes,
 
\begin{equation} E(\delta X \delta Y) = N_{c+f}^{late}+N_{f}^{late}\end{equation}
  In this case, eq. (\ref{frac_err}) becomes,
\begin{equation} 
    E(\delta f_{late}) =  f_{late}\times \sqrt{\frac{ N_{c+f}^{late}+N_{f}^{late}}{(  N_{c+f}^{late}-N_{f}^{late}  )^2}  +    \frac{N_{c+f}^{all}+N_{f}^{all}}{(N_{c+f}^{all}-N_{f}^{all})^2}   - \frac{2(N_{c+f}^{late}+N_{f}^{late})}{(N_{c+f}^{late}-N_f^{late})(N_{c+f}^{all}-N_f^{all})} }
\label{frac_err_referee}
 \end{equation}

\newpage

\newpage

\newpage

\newpage

\newpage

\newpage

\newpage

\newpage

\newpage
\newpage

%
%
%
%

\newpage

\newpage

%
%

\newpage

\newpage

\newpage

\newpage

\newpage

%
%
%

%

\begin{table}[h]
\caption{
\label{tab:correlation}
  Spearman's correlation coefficients between $z$ and fractions of late type
 galaxeis. 514 clusters with richness$>$25 are chosen as a sample.
}
\begin{center}
\begin{tabular}{llll}
\hline
  & Correlation coefficient  & Significance  & N clusters \\
\hline
\hline
 $f_b$      &        0.238 & 4.4$\times$10$^{-8}$ &         514\\ 
 $f_{u-r}$  &        0.234 & 7.6$\times$10$^{-8}$   &      514\\ 
 $f_{exp}$  &       0.194 & 9.6$\times$10$^{-6}$ &         514\\
 $f_{Cin}$  &       0.223 & 2.9$\times$10$^{-7}$  &         514\\ 
\hline
\end{tabular}
\end{center}
\end{table}

\begin{table}[h]
\caption{
\label{tab:kstest}
 Significances in Kolomogorov-Smirnov tests between distributions
 for z$\leq$0.15 and 0.15$<$z$<$0.3. In all cases, Kolomogorov-Smirnov tests show the distributions for
 the lower redshift sample and the higher redshift sample are significantly different. 
}
\begin{center}
\begin{tabular}{ll}
\hline
  & Significance  \\
\hline
\hline
 $f_b$      &  2.9$\times$10$^{-3}$ \\ 
 $f_{u-r}$  &  1.0$\times$10$^{-3}$ \\ 
 $f_{exp}$  &  3.4$\times$10$^{-4}$ \\
 $f_{Cin}$  &  2.9$\times$10$^{-3}$ \\ 
\hline
\end{tabular}
\end{center}
\end{table}

\begin{table}[h]
\caption{
\label{tab:error_comparison}
Scatters in late-type fractions around the best-fit line are compared
 with median errors of late-type fraction calculated with equation (\ref{frac_err}) 
}
\begin{center}
\begin{tabular}{lllll}
\hline
     &  $f_b$  & $f_{u-r}$ & $f_{exp}$ & $f_{Cin}$\\
\hline
\hline
 Real scatter (1$\sigma$) &  0.169  & 0.183  & 0.171 & 0.163    \\
 Error estimate           &  0.078  & 0.069  & 0.050 & 0.089    \\
\hline
\end{tabular}
\end{center}
\end{table}

\begin{table}[h]
\caption{
\label{tab:cin_g_r}
Change in the fraction of galaxies with $C_{in}>$0.4 (late type) in two different filters($g,r$).
}
\begin{center}
\begin{tabular}{llll}
\hline
band & N($C_{in}>$0.4)  & N(total) & Percentage(\%) \\
\hline
\hline
 $g$ & 802 & 1336 &     60.0       \\
 $r$ & 787 & 1336 &     58.9       \\
\hline
 Difference & 15  & 1336   &  1.1\\ 
\hline
\end{tabular}
\end{center}
\end{table}

\begin{table}[h]
\caption{
\label{tab:exp_g_r}
Change in the fraction of galaxies with exponential fit likelihood
 greater than de Vaucouleur likelihood (late type) in two different
 filters($g,r$). Since we discard the galaxies
 with the same likelihood in this analysis, the total number of galaxies
 in the sample are different in $g$ and $r$.
}
\begin{center}
\begin{tabular}{llll}
\hline
band & N(late)  & N(late + early) & Percentage(\%) \\
\hline
\hline
 $g$ & 503 & 804 &     62.6       \\
 $r$ & 476 & 792 &     60.1       \\
\hline
 Difference & -  & -   &  2.5\\ 
\hline
\end{tabular}
\end{center}
\end{table}


\begin{thebibliography}{DUM}\label{refrence}

\bibitem[Abadi, Moore, \& Bower(1999)]{1999MNRAS.308..947A} Abadi, M.~G., 
Moore, B., \& Bower, R.~G.\ 1999, \mnras, 308, 947 

\bibitem{Abell58}         Abell, G. 1958,         APJS, 3, 211

\bibitem{Abell89}         Abell, G., Corwin, H. \& Olowin, R. 1989, APJS, 70, 1

\bibitem[Abraham et al.(1996)]{1996ApJS..107....1A} Abraham, R.~G., van den 
Bergh, S., Glazebrook, K., Ellis, R.~S., Santiago, B.~X., Surma, P., \& Griffiths, R.~E.\ 1996, \apjs, 107, 1 


\bibitem[Abraham, Valdes, Yee, \& van den Bergh(1994)]{1994ApJ...432...75A} 
Abraham, R.~G., Valdes, F., Yee, H.~K.~C., \& van den Bergh, S.\ 1994, 
\apj, 432, 75 

\bibitem[Abraham et al.(1996)]{1996ApJ...471..694A} Abraham, R.~G.~et al.\ 
1996, \apj, 471, 694 


\bibitem[Abraham \& van den Bergh(2001)]{2001Sci...293.1273A} Abraham, 
R.~G.~\& van den Bergh, S.\ 2001, Science, 293, 1273 

\bibitem[Allen \& Fabian(1998)]{1998MNRAS.297L..57A} Allen, S.~W.~\& 
Fabian, A.~C.\ 1998, \mnras, 297, L57 




\bibitem[Allington-Smith, Ellis, Zirbel, \& Oemler(1993)]{1993ApJ...404..521A} Allington-Smith, J.~R., Ellis, R., 
Zirbel, E.~L., \& Oemler, A.~J.\ 1993, \apj, 404, 521 


\bibitem[Andreon \& Ettori(1999)]{1999ApJ...516..647A} Andreon, S.~\& Ettori, S.\ 1999, \apj, 516, 647 


 \bibitem{}Annis, J. et al. 2002 $in$ $preparation$ 

\bibitem[Arnaud \& Evrard(1999)]{1999MNRAS.305..631A} Arnaud, M.~\& Evrard, 
A.~E.\ 1999, \mnras, 305, 631 



\bibitem[Bahcall(1974)]{1974ApJ...193..529B} Bahcall, N.~A.\ 1974, \apj, 
193, 529 

\bibitem[Bahcall(1977)]{1977ApJ...218L..93B} Bahcall, N.~A.\ 1977, \apjl, 
218, L93 

 \bibitem{}Bahcall, N. A. et al. 2002 $in$ $preparation$ 

\bibitem[Balogh et al.(1997)]{1997ApJ...488L..75B} Balogh, M.~L., Morris, 
S.~L., Yee, H.~K.~C., Carlberg, R.~G., \& Ellingson, E.\ 1997, \apjl, 488, 
L75 



\bibitem[Balogh et al.(1999)]{1999ApJ...527...54B} Balogh, M.~L., Morris, 
S.~L., Yee, H.~K.~C., Carlberg, R.~G., \& Ellingson, E.\ 1999, \apj, 527, 54 

\bibitem[Balogh, Navarro, \& Morris(2000)]{2000ApJ...540..113B} Balogh, 
M.~L., Navarro, J.~F., \& Morris, S.~L.\ 2000, \apj, 540, 113 


\bibitem[Balogh, Christlein, Zabludoff, \& Zaritsky(2001)]{2001ApJ...557..117B} Balogh, M.~L., Christlein, D., 
Zabludoff, A.~I., \& Zaritsky, D.\ 2001, \apj, 557, 117 



\bibitem[Balogh et al.(2002)]{2002MNRAS.337..256B} Balogh, M., Bower,
R.~G., Smail, I., Ziegler, B.~L., Davies, R.~L., Gaztelu, A., \& Fritz, A.\
2002, \mnras, 337, 256



\bibitem[Bekki(1998)]{1998ApJ...502L.133B} Bekki, K.\ 1998, \apjl, 502, 
L133 


\bibitem[Bekki, Shioya, \& Couch(2001)]{2001ApJ...547L..17B} Bekki, K., 
Shioya, Y., \& Couch, W.~J.\ 2001, \apjl, 547, L17 


\bibitem[Bekki, Couch, \& Shioya(2002)]{2002ApJ...577..651B} Bekki, K.,
Couch, W.~J., \& Shioya, Y.\ 2002, \apj, 577, 651

\bibitem[Binggeli, Sandage, \& Tammann(1988)]{1988ARA&A..26..509B} 
Binggeli, B., Sandage, A., \& Tammann, G.~A.\ 1988, \araa, 26, 509 

 \bibitem{}Binney, J., \& Tremaine, S. 1987, Galactic Dynamics
				      (Princeton: Princeton Univ. Press)

\bibitem[Blanton et al.(2001)]{2001AJ....121.2358B} Blanton, M.~R.~et al.\ 
2001, \aj, 121, 2358 

\bibitem[Blanton et al.(2002)]{2002AJ} Blanton, M.R., Lupton, R.H.,
				      Maley, F.M., Young, N., Zehavi,
				      I., and  Loveday, J. 2002, AJ,
				      submitted

\bibitem[Bower et al.(1994)]{1994MNRAS.268..345B} Bower, R.~G., Bohringer, 
H., Briel, U.~G., Ellis, R.~S., Castander, F.~J., \& Couch, W.~J.\ 1994, 
\mnras, 268, 345 


\bibitem[Boyce et al.(2001)]{2001MNRAS.328..277B} Boyce, P.~J., Phillipps, 
S., Jones, J.~B., Driver, S.~P., Smith, R.~M., \& Couch, W.~J.\ 2001, 
\mnras, 328, 277. 

\bibitem[Bothun \& Dressler(1986)]{1986ApJ...301...57B} Bothun, G.~D.~\& Dressler, A.\ 1986, \apj, 301, 57 


\bibitem[Butcher \& Oemler(1978)]{1978ApJ...226..559B} Butcher, H.~\& Oemler, A.\ 1978, \apj, 226, 559 


\bibitem[Butcher \& Oemler(1984)]{1984ApJ...285..426B} Butcher, H.~\& Oemler, A.\ 1984, \apj, 285, 426 

\bibitem[Byrd \& Valtonen(1990)]{1990ApJ...350...89B} Byrd, G.~\& Valtonen, 
M.\ 1990, \apj, 350, 89

\bibitem[Cavaliere, Menci, \& Tozzi(1997)]{1997ApJ...484L..21C} Cavaliere, 
A., Menci, N., \& Tozzi, P.\ 1997, \apjl, 484, L21 


\bibitem[Couch, Ellis, Sharples, \& Smail(1994)]{1994ApJ...430..121C} 
Couch, W.~J., Ellis, R.~S., Sharples, R.~M., \& Smail, I.\ 1994, \apj, 430, 
121 


\bibitem[Couch et al.(1998)]{1998ApJ...497..188C} Couch, W.~J., Barger, 
A.~J., Smail, I., Ellis, R.~S., \& Sharples, R.~M.\ 1998, \apj, 497, 188 


\bibitem[Cowie, Songaila, and Barger(1999)]{1999AJ....118..603C} Cowie, 
L.~L., Songaila, A., \& Barger, A.~J.\ 1999, \aj, 118, 603 



\bibitem[David et al.(1993)]{1993ApJ...412..479D} David, L.~P., Slyz, A., 
Jones, C., Forman, W., Vrtilek, S.~D., \& Arnaud, K.~A.\ 1993, \apj, 412, 
479 



\bibitem[Diaferio et al. (2001)]{} Diaferio, A., Kauffmann, G., Balogh, M.L., White, S.D.M., Schade, D. \& Ellingson, E. 2001, \mnras, 323, 999.

 \bibitem[Doi, Fukugita, \& Okamura(1993)]{1993MNRAS.264..832D} Doi, M., 
Fukugita, M., \& Okamura, S.\ 1993, \mnras, 264, 832 


\bibitem[Dressler(1980)]{1980ApJ...236..351D} Dressler, A.\ 1980, \apj, 
236, 351 

\bibitem[Dressler \& Gunn(1992)]{1992ApJS...78....1D} Dressler, A.~\& Gunn, 
J.~E.\ 1992, \apjs, 78, 1 


\bibitem[Dressler et al.(1997)]{1997ApJ...490..577D} Dressler, A.~et al.\ 
1997, \apj, 490, 577 


\bibitem[Dressler et al.(1999)]{1999ApJS..122...51D} Dressler, A., Smail, 
I., Poggianti, B.~M., Butcher, H., Couch, W.~J., Ellis, R.~S., \& Oemler, 
A.~J.\ 1999, \apjs, 122, 51 


\bibitem[Edge \& Stewart(1991)]{1991MNRAS.252..414E} Edge, A.~C.~\& 
Stewart, G.~C.\ 1991, \mnras, 252, 414 



\bibitem[Eisenstein et al.(2001)]{2001AJ....122.2267E} Eisenstein, D.~J.~et 
al.\ 2001, \aj, 122, 2267 

\bibitem[Ellingson, Lin, Yee, \& Carlberg(2001)]{2001ApJ...547..609E} 
Ellingson, E., Lin, H., Yee, H.~K.~C., \& Carlberg, R.~G.\ 2001, \apj, 547, 
609 

\bibitem[Evrard(1991)]{1991MNRAS.248P...8E} Evrard, A.~E.\ 1991, \mnras, 
248, 8P 


\bibitem[Fairley et al.(2002)]{2002MNRAS.330..755F} Fairley, B.~W., Jones, 
L.~R., Wake, D.~A., Collins, C.~A., Burke, D.~J., Nichol, R.~C., \& Romer, 
A.~K.\ 2002, \mnras, 330, 755 


\bibitem[Farouki \& Shapiro(1980)]{1980ApJ...241..928F} Farouki, R.~\& Shapiro, S.~L.\ 1980, \apj, 241, 928 


\bibitem[Fasano et al.(2000)]{2000ApJ...542..673F} Fasano, G.,
				      Poggianti, B.~M., Couch, W.~J.,
				      Bettoni, D., Kj{\ae}rgaard, P., \&
				      Moles, M. 2000, \apj, 542, 673 


 \bibitem{}Finoguenov, A., Briel, U.G., Henry, J.P., 2003, submitted to
				      A\&A


\bibitem[Fujita(1998)]{1998ApJ...509..587F} Fujita, Y.\ 1998, \apj, 509, 
587 

\bibitem[Fujita \& Nagashima(1999)]{1999ApJ...516..619F} Fujita, Y.~\& Nagashima, M.\ 1999, \apj, 516, 619 

\bibitem[Fujita(2001)]{2001ApJ...550..612F} Fujita, Y.\ 2001, \apj, 550, 
612 


\bibitem[Fukugita, Shimasaku, \& Ichikawa(1995)]{1995PASP..107..945F}
			    Fukugita, M., Shimasaku, K., \& Ichikawa,
			    T.\ 1995, \pasp, 107, 945

\bibitem[Fukugita et al.(1996)]{1996AJ....111.1748F} Fukugita, M., 
Ichikawa, T., Gunn, J.~E., Doi, M., Shimasaku, K., \& Schneider, D.~P.\ 
1996, \aj, 111, 1748. 


\bibitem[Gal et al.(2000)]{2000AJ....120..540G} Gal, R.~R., de Carvalho, 
R.~R., Brunner, R., Odewahn, S.~C., \& Djorgovski, S.~G.\ 2000, \aj, 120, 
540 

\bibitem[Garilli et al.(1996)]{1996ApJS..105..191G} Garilli, B., Bottini, 
D., Maccagni, D., Carrasco, L., \& Recillas, E.\ 1996, \apjs, 105, 191 

\bibitem[Goto et al.(2002)]{2002AJ....123.1807G} Goto, T.~et al.\ 2002a, 
\aj, 123, 1807. 

\bibitem{}Goto, T., Okamura, S. \& Brinkman, J. 2002b, PASJ, 54, 4 

\bibitem[Gladders \& Yee(2000)]{2000AJ....120.2148G} Gladders, M.~D.~\& Yee, H.~K.~C.\ 2000, \aj, 120, 2148 


\bibitem[Gunn \& Gott(1972)]{1972ApJ...176....1G} Gunn, J.~E.~\& Gott, 
J.~R.~I.\ 1972, \apj, 176, 1 


 \bibitem{}Gunn, J.E., Carr, M., Rockosi, C., Sekiguchi, M., Berry, K., Elms, B.,	de Haas, E., Ivezic , Z. et al.\ 1998, AJ, 116, 3040 

\bibitem{}Gunn, J.E., Hoessel, J.G., \& Oke, J.B.\ 1986, \apj, 306, 30.


\bibitem[Hammer et al.(1997)]{1997ApJ...481...49H} Hammer, F., Flores, H., 
Lilly, S.~J., Crampton, D., Le Fevre, O., Rola, C., Mallen-Ornelas, G., 
Schade, D., \& Tresse, L.\ 1997, \apj, 481, 49 


\bibitem[Henry \& Arnaud(1991)]{1991ApJ...372..410H} Henry, J.~P.~\& 
Arnaud, K.~A.\ 1991, \apj, 372, 410 



\bibitem[Hogg, Finkbeiner, Schlegel, \& Gunn(2001)]{2001AJ....122.2129H} 
Hogg, D.~W., Finkbeiner, D.~P., Schlegel, D.~J., \& Gunn, J.~E.\ 2001, \aj, 
122, 2129 


\bibitem[Icke(1985)]{1985A&A...144..115I} Icke, V.\ 1985, \aap, 144, 115 

\bibitem[Jones \& Forman(1978)]{1978ApJ...224....1J} Jones, C.~\& Forman, 
W.\ 1978, \apj, 224, 1 

\bibitem[Jones \& Forman(1999)]{1999ApJ...511...65J} Jones, C.~\& Forman, 
W.\ 1999, \apj, 511, 65 




\bibitem[Kauffmann(1995)]{1995MNRAS.274..153K} Kauffmann, G.\ 1995, \mnras, 
274, 153 


\bibitem[Kent(1981)]{1981ApJ...245..805K} Kent, S.~M.\ 1981, \apj, 245, 805 

\bibitem[Kim et al.(2002)]{2002AJ....123...20K} Kim, R.~S.~J.~et al.\ 2002, 
\aj, 123, 20 


\bibitem[Kodama et al.(2001)]{2001ApJ...562L...9K} Kodama, T., Smail, I., 
Nakata, F., Okamura, S., \& Bower, R.~G.\ 2001, \apjl, 562, L9 


\bibitem[Kodama \& Bower(2001)]{2001MNRAS.321...18K} Kodama, T.~\& Bower, 
R.~G.\ 2001, \mnras, 321, 18 

\bibitem[Kodama \& Smail(2001)]{2001MNRAS.326..637K} Kodama, T.~\& Smail, 
I.\ 2001, \mnras, 326, 637 

\bibitem[Lahav et al.(1995)]{1995Sci...267..859L} Lahav, O.~et al.\ 1995, 
Science, 267, 859 


\bibitem[Larson, Tinsley, \& Caldwell(1980)]{1980ApJ...237..692L} Larson, 
R.~B., Tinsley, B.~M., \& Caldwell, C.~N.\ 1980, \apj, 237, 692 

\bibitem[Lavery \& Henry(1988)]{1988ApJ...330..596L} Lavery, R.~J.~\&
				      Henry, J.~P.\ 1988, \apj, 330, 596 

\bibitem[Lea \& Henry(1988)]{1988ApJ...332...81L} Lea, S.~M.~\& Henry, 
J.~P.\ 1988, \apj, 332, 81 


\bibitem[Lilly, Le Fevre, Hammer, and Crampton(1996)]{1996ApJ...460L...1L} 
Lilly, S.~J., Le Fevre, O., Hammer, F., \& Crampton, D.\ 1996, \apjl, 460, 
L1 



\bibitem[Lilly et al.(1998)]{1998ApJ...500...75L} Lilly, S., Schade, D., 
Ellis, R., Le Fevre, O., Brinchmann, J., Tresse, L., Abraham, R., Hammer, 
F., et al.\ 1998, \apj, 500, 75 



\bibitem[Lin et al.(1999)]{1999ApJ...518..533L} Lin, H., Yee, H.~K.~C., 
Carlberg, R.~G., Morris, S.~L., Sawicki, M., Patton, D.~R., Wirth, G., \& Shepherd, C.~W.\ 1999, \apj, 518, 533 


\bibitem[Loveday, Peterson, Efstathiou, \& Maddox(1992)]{1992ApJ...390..338L} Loveday, J., Peterson, B.~A., 
Efstathiou, G., \& Maddox, S.~J.\ 1992, \apj, 390, 338 


\bibitem[Lupton, Gunn, \& Szalay(1999)]{1999AJ....118.1406L} Lupton, R.~H., Gunn, J.~E., \& Szalay, A.~S.\ 1999, \aj, 118, 1406

\bibitem[Lupton et al.(2001)]{2001adass..10..269L} Lupton, R.~H., Gunn, J.~E., Ivezi{\' c}, Z., Knapp, G.~R., Kent, S., \& Yasuda, N.\ 2001, Astronomical Data Analysis Software and Systems X, ASP Conference Proceedings, Vol.~238.~Edited by F.~R.~Harnden, Jr., Francis A.~Primini, and Harry E.~Payne.~San Francisco: Astronomical Society of the Pacific, ISSN: 1080-7926, 2001., p.269, 10, 269

 \bibitem{}Lupton, R. et al. 2002, $in$ $preparation$

\bibitem[Madau et al.(1996)]{1996MNRAS.283.1388M} Madau, P., Ferguson, 
H.~C., Dickinson, M.~E., Giavalisco, M., Steidel, C.~C., \& Fruchter, A.\ 
1996, \mnras, 283, 1388 


\bibitem[Makino \& Hut(1997)]{1997ApJ...481...83M} Makino, J.~\& Hut, P.\ 
1997, \apj, 481, 83 


\bibitem[Mamon(1992)]{1992ApJ...401L...3M} Mamon, G.~A.\ 1992, \apjl, 401, 
L3 


\bibitem[Margoniner \& de Carvalho(2000)]{2000AJ....119.1562M} Margoniner, 
V.~E.~\& de Carvalho, R.~R.\ 2000, \aj, 119, 1562 


\bibitem[Margoniner, de Carvalho, Gal, \& Djorgovski(2001)]{2001ApJ...548L.143M} Margoniner, V.~E., de Carvalho, 
R.~R., Gal, R.~R., \& Djorgovski, S.~G.\ 2001, \apjl, 548, L143 


\bibitem[Markevitch(1998)]{1998ApJ...504...27M} Markevitch, M.\ 1998, \apj, 
504, 27 



\bibitem[Marzke, Geller, Huchra, \& Corwin(1994)]{1994AJ....108..437M} 
Marzke, R.~O., Geller, M.~J., Huchra, J.~P., \& Corwin, H.~G.\ 1994, \aj, 
108, 437 

\bibitem[Marzke \& da Costa(1997)]{1997AJ....113..185M} Marzke, R.~O.~\& da 
Costa, L.~N.\ 1997, \aj, 113, 185 


\bibitem[Marzke et al.(1998)]{1998ApJ...503..617M} Marzke, R.~O., da Costa, 
L.~N., Pellegrini, P.~S., Willmer, C.~N.~A., \& Geller, M.~J.\ 1998, \apj, 
503, 617 



\bibitem[Metevier, Romer, \& Ulmer(2000)]{2000AJ....119.1090M} Metevier, 
A.~J., Romer, A.~K., \& Ulmer, M.~P.\ 2000, \aj, 119, 1090 

\bibitem[Mihos(1995)]{1995ApJ...438L..75M} Mihos, J.~C.\ 1995, \apjl, 438, 
L75 


 \bibitem{}Miller, C. J. et al. astro-ph/9912362
 \bibitem{}Miller, C. J. et al. 2002 $in$ $preparation$ 

 \bibitem[Mitchell, Ives, and Culhane(1977)]{1977MNRAS.181P..25M} Mitchell, 
R.~J., Ives, J.~C., \& Culhane, J.~L.\ 1977, \mnras, 181, 25P 

 \bibitem[Mo \& Mao(2002)]{2002MNRAS.333..768M} Mo, H.~J.~\& Mao, S.\
				      2002, \mnras, 333, 768


\bibitem[Moore et al.(1996)]{1996Natur.379..613M} Moore, B., Katz, N., 
Lake, G., Dressler, A., \& Oemler, A.\ 1996, \nat, 379, 613 

\bibitem[Moore, Lake, Quinn, \& Stadel(1999)]{1999MNRAS.304..465M} Moore, 
B., Lake, G., Quinn, T., \& Stadel, J.\ 1999, \mnras, 304, 465 


\bibitem[Morgan(1958)]{1958PASP...70..364M} Morgan, W.~W.\ 1958, \pasp, 70, 
364 

\bibitem[Morgan(1959)]{1959PASP...71..394M} Morgan, W.~W.\ 1959, \pasp, 71, 
394 

\bibitem[Newberry, Kirshner, \& Boroson(1988)]{1988ApJ...335..629N} 
Newberry, M.~V., Kirshner, R.~P., \& Boroson, T.~A.\ 1988, \apj, 335,
				      629 

\bibitem{}Oh, S.P. \& Benson, A.J. 2003, submitted to MNRAS 

\bibitem[Ostriker(1980)]{1980ComAp...8..177O} Ostriker, J.~P.\ 1980,
Comments on Astrophysics, 8, 177

\bibitem[Pier et al.(2003)]{2003AJ....125.1559P} Pier, J.~R., Munn, J.~A.,
Hindsley, R.~B., Hennessy, G.~S., Kent, S.~M., Lupton, R.~H., \& Ivezi{\'
c}, {\v Z}.\ 2003, \aj, 125, 1559


\bibitem[Poggianti et al.(1999)]{1999ApJ...518..576P} Poggianti, B.~M., 
Smail, I., Dressler, A., Couch, W.~J., Barger, A.~J., Butcher, H., Ellis, 
R.~S., \& Oemler, A.~J.\ 1999, \apj, 518, 576 

\bibitem[Quilis, Moore, \& Bower(2000)]{2000Sci...288.1617Q} Quilis, V., 
Moore, B., \& Bower, R.\ 2000, Science, 288, 1617 


\bibitem[Rakos \& Schombert(1995)]{1995ApJ...439...47R} Rakos, K.~D.~\&
				      Schombert, J.~M.\ 1995, \apj, 439,
				      47 

\bibitem[Reichart, Castander, \& Nichol(1999)]{1999ApJ...516....1R} 
Reichart, D.~E., Castander, F.~J., \& Nichol, R.~C.\ 1999, \apj, 516, 1 


\bibitem[Schade et al.(1996)]{1996ApJ...464...79S} Schade, D., Lilly, 
S.~J., Le Fevre, O., Hammer, F., \& Crampton, D.\ 1996, \apj, 464, 79 


\bibitem[Schlegel, Finkbeiner, \& Davis(1998)]{1998ApJ...500..525S} 
Schlegel, D.~J., Finkbeiner, D.~P., \& Davis, M.\ 1998, \apj, 500, 525 

 \bibitem{}Scranton, R.  et al. 2002 $in$ $preparation$

\bibitem[Sellwood \& Carlberg(1984)]{1984ApJ...282...61S} Sellwood, 
J.~A.~\& Carlberg, R.~G.\ 1984, \apj, 282, 61 


\bibitem[Shimasaku et al.(2001)]{2001AJ....122.1238S} Shimasaku, K.~et
	 al.\ 2001, \aj, 122, 1238
\bibitem[Shioya, Bekki, Couch, \& De Propris(2002)]{2002ApJ...565..223S} 
Shioya, Y., Bekki, K., Couch, W.~J., \& De Propris, R.\ 2002, \apj, 565, 223 

\bibitem[Smail et al.(1997)]{1997ApJS..110..213S} Smail, I., Dressler, A., 
Couch, W.~J., Ellis, R.~S., Oemler, A.~J., Butcher, H., \& Sharples, R.~M.\ 
1997, \apjs, 110, 213 

\bibitem[Smail, Edge, Ellis, \& Blandford(1998)]{1998MNRAS.293..124S} 
Smail, I., Edge, A.~C., Ellis, R.~S., \& Blandford, R.~D.\ 1998, \mnras, 
293, 124 


\bibitem[Smith et al.(2002)]{2002AJ....123.2121S} Smith, J.~A.~et al.\ 
2002, \aj, 123, 2121 

\bibitem[Spitzer \& Baade(1951)]{1951ApJ...113..413S} Spitzer, L.~J.~\& Baade, W.\ 1951, \apj, 113, 413 

\bibitem[Stevens, Acreman, \& Ponman(1999)]{1999MNRAS.310..663S} Stevens, 
I.~R., Acreman, D.~M., \& Ponman, T.~J.\ 1999, \mnras, 310, 663 


\bibitem[Strauss et al.(2002)]{2002AJ....124.1810S} Strauss, M.~A.~et al.\
2002, \aj, 124, 1810

\bibitem[Strateva et al.(2001)]{2001AJ....122.1861S} Strateva, I.~et	 al.\ 2001, \aj, 122, 1861


\bibitem[Stoughton et al.(2002)]{2002AJ....123..485S} Stoughton, C.~et al.\ 
2002, \aj, 123, 485 


\bibitem[Sullivan et al.(2000)]{2000MNRAS.312..442S} Sullivan, M., Treyer, 
M.~A., Ellis, R.~S., Bridges, T.~J., Milliard, B., \& Donas, J.~; 2000, 
\mnras, 312, 442 




\bibitem[Treyer et al.(1998)]{1998MNRAS.300..303T} Treyer, M.~A., Ellis, 
R.~S., Milliard, B., Donas, J., \& Bridges, T.~J.\ 1998, \mnras, 300, 303 






\bibitem[Valotto, Moore, \& Lambas(2001)]{2001ApJ...546..157V} Valotto, 
C.~A., Moore, B., \& Lambas, D.~G.\ 2001, \apj, 546, 157 


\bibitem[Valluri(1993)]{1993ApJ...408...57V} Valluri, M.\ 1993, \apj, 408, 
57 


\bibitem[van den Bergh(1976)]{1976ApJ...206..883V} van den Bergh, S.\ 1976,
\apj, 206, 883




\bibitem[Vollmer, Cayatte, Balkowski, \& Duschl(2001)]{2001ApJ...561..708V} 
Vollmer, B., Cayatte, V., Balkowski, C., \& Duschl, W.~J.\ 2001, \apj, 561, 
708 


\bibitem[Wang \& Ulmer(1997)]{1997MNRAS.292..920W} Wang, Q.~D.~\& Ulmer, 
M.~P.\ 1997, \mnras, 292, 920 


\bibitem[White, Jones, and Forman(1997)]{1997MNRAS.292..419W} White, D.~A., 
Jones, C., \& Forman, W.\ 1997, \mnras, 292, 419 

\bibitem[Wilson, Cowie, Barger, \& Burke(2002)]{2002AJ....124.1258W}
Wilson, G., Cowie, L.~L., Barger, A.~J., \& Burke, D.~J.\ 2002, \aj, 124,
1258


\bibitem[Wu, Xue, and Fang(1999)]{1999ApJ...524...22W} Wu, X., Xue, Y., \& 
Fang, L.\ 1999, \apj, 524, 22 


\bibitem[Xue \& Wu(2000)]{2000ApJ...538...65X} Xue, Y.~\& Wu, X.\ 2000, 
\apj, 538, 65 


\bibitem[Yagi et al.(2002)]{2002AJ....123...87Y} Yagi, M., Kashikawa, N., 
Sekiguchi, M., Doi, M., Yasuda, N., Shimasaku, K., \& Okamura, S.\ 2002, 
\aj, 123, 87 

\bibitem[Yagi et al.(2002)]{2002AJ....123...66Y} Yagi, M., Kashikawa, N., 
Sekiguchi, M., Doi, M., Yasuda, N., Shimasaku, K., \& Okamura, S.\ 2002, 
\aj, 123, 66 


\bibitem[Yee, Ellingson, \& Carlberg(1996)]{1996ApJS..102..269Y} Yee, 
H.~K.~C., Ellingson, E., \& Carlberg, R.~G.\ 1996, \apjs, 102, 269 


\bibitem[York et al.(2000)]{2000AJ....120.1579Y} York, D.~G.~et al.\ 2000, \aj, 120, 1579




\end{thebibliography}
\end{document}